\documentclass[10pt]{iopart}
\usepackage{iopams,color,fullpage}
\usepackage[colorlinks=true,linkcolor=blue,citecolor=red,bookmarksnumbered=true]{hyperref}
\usepackage{amssymb,times,fourier,charter}

\advance\textwidth -8mm
\advance\hoffset 4mm
\advance\textheight -4mm
\advance\voffset 2mm

\makeatletter

\eqnobysec
\def\eqref#1{(\ref{#1})}

\def\numparts{\refstepcounter{equation}%
     \setcounter{eqnval}{\value{equation}}%
     \setcounter{equation}{0}%
     \def\theequation{\arabic{section}.\arabic{eqnval}{\it\alph{equation}}}}
\def\endnumparts{\def\theequation{\arabic{section}.\arabic{equation}}%
     \setcounter{equation}{\value{eqnval}}}
\def\bse{\numparts}
\def\ese{\endnumparts}
\def\bea{\begin{eqnarray}}
\def\eea{\end{eqnarray}}
\def\be{\begin{equation}}
\def\ee{\end{equation}}

\definecolor{darkblue}{rgb}{0,0,0.8}

\renewcommand\section{\@startsection {section}{1}{\z@}%
  {-3.2ex \@plus -1ex \@minus -.2ex}%
  {2ex \@plus.2ex}%
  {\color{darkblue}\reset@font\normalsize\bfseries\raggedright}}
\renewcommand\subsection{\@startsection{subsection}{2}{\z@}%
  {-3ex\@plus -1ex \@minus -.2ex}%
  {1ex \@plus .2ex}%
  {\color{darkblue}\reset@font\normalsize\itshape\raggedright}}

\newcommand\partialderiv[3][]{\frac{\partial^{#1}#2}{\partial {#3}^{#1}}}
\let\true@epsilon=\epsilon
\let\epsilon=\varepsilon
\let\^=\hat
\let\==\bar
\let\~=\tilde
\let\@=\mathbf
\let\_=\mathsf

\let\.=\dot
\def\fbf#1{\setbox0=\hbox{$#1$}\kern-0.10\wd0
  \lower0.02em\copy0\kern-\wd0 \lower0.02em\hbox{\kern+0.05em\copy0}\kern-\wd0
  \raise0.02em\copy0\kern-\wd0 \raise0.02em\hbox{\kern-0.05em\box0}}
  
\def\d{{\mathrm{d}}}
\def\e{{\mathrm{e}}}

\def\Ones{{\mathbb{1}}}

\def\Real{\mathbb{R}}

\def\sn{\mathop{\rm sn}\nolimits}

\def\diag{\mathop{\rm diag}\nolimits}

\def\half{{\textstyle\frac12}}

\def\pvint{\int\kern-0.94em-\kern0.2em}

\let\le=\leqslant
\let\ge=\geqslant
\def\E{E_m}
\def\K{K_m}
\def\txtfrac#1#2{{\textstyle\frac{#1}{#2}}}
\def\deriv#1#2{{\frac{d#1}{d#2}}}
\def\partialderiv#1#2{{\frac{\partial #1}{\partial #2}}}
\def\diag{\mathop{\rm diag}\nolimits}
\let\truenabla=\nabla
\def\bfnabla{{\fbf\truenabla}}
\makeatother

\def\Dy#1{\frac{D#1}{Dy}}
\def\Dz#1{\frac{D#1}{Dz}}

\begin{document}
\title{\color{red}%
Whitham modulation theory for the defocusing nonlinear Schr\"odinger
equation in two and three spatial dimensions}
\author{Asela Abeya$^1$, Gino Biondini$^1$ and Mark A. Hoefer$^2$}
\address{$^1$~State University of New York, Department of Mathematics, Buffalo, NY 14260, USA}
\address{$^2$~University of Colorado, Department of Applied Mathematics, Boulder, CO 80309, USA} 
\date{\today}
\def\submitto#1{\relax}

\begin{abstract}
The Whitham modulation equations for the defocusing nonlinear Schr\"odinger (NLS) equation in two, three and higher spatial dimensions are
derived using a two-phase ansatz for the periodic traveling wave solutions and by period-averaging the conservation laws of the NLS
equation.  
The resulting Whitham modulation equations are written in vector form, which allows one to show that they preserve the rotational invariance of the NLS equation, as well as the invariance with respect to scaling and Galilean transformations, and
to immediately generalize the calculations from two spatial dimensions to three.  
The transformation to Riemann-type variables is described in detail; 
the harmonic and soliton limits of the Whitham modulation equations are explicitly written down; 
and the reduction of the Whitham equations to those for the radial NLS equation is explicitly carried out.  
Finally, the extension of the theory to higher spatial dimensions is briefly outlined.  
The multidimensional NLS-Whitham equations obtained here may be used to study large amplitude wavetrains in a variety of applications including nonlinear photonics and matter waves.
\par\medskip
\noindent
\today
\end{abstract}

\section{Introduction} 
\label{s:intro}

The nonlinear Schr\"odinger (NLS) equation in one, two and three
spatial dimensions is a ubiquitous model in nonlinear science.
One
reason is its universality as a model for the evolution of weakly
nonlinear dispersive wave trains
\cite{BenneyNewell,CalogeroEckhaus,SulemSulem}.
The NLS equation
arises as the governing equation in a broad variety of physical
contexts, ranging from water waves to optics, acoustics, Bose-Einstein
condensates and beyond
\cite{AS1981,InfeldRowlands,KevrekidisFrantzeskakisCarretero2009,NewellMoloney}.
As a result, enormous attention has been devoted over the last half
century to the study of its solutions.
It is also the case that in many physical situations, dispersive
effects are much weaker than nonlinear ones and these scenarios, which
can be formulated as small dispersion limits of the governing
equations, give rise to a variety of interesting physical phenomena
\cite{ElHoefer}.
In particular, the small dispersion limits often
lead to the formation of dispersive shock waves, a coherent, slowly
modulated and expanding train of nonlinear oscillations.

A powerful tool in the study of small dispersion limits is Whitham
modulation theory (also simply called Whitham theory) \cite{Whitham1965,Whitham1974}.
Whitham theory is an asymptotic framework within which one can derive
the Whitham modulation equations or Whitham equations for brevity.
The Whitham equations are a system of first-order, quasi-linear
partial differential equations (PDEs) that govern the evolution of the
periodic traveling wave solutions of the original PDE over spatial and
temporal scales that are larger than the traveling wave
solution's wavelength and period, respectively.
Whitham theory does
not require integrability of the original PDE, and therefore it can
also be applied to non-integrable PDEs.
Thanks to Whitham theory and, when applicable, the inverse scattering
transform (IST), much is known about small dispersion limits for
(1+1)-dimensional nonlinear wave equations 
(e.g., see \cite{PhysD333p1,newref4,ElHoefer,newref3,k2000,newref1} and references therein).
On the other
hand, small dispersion limits for (2+1)-dimensional systems have been
much less studied and (3+1)-dimensional systems apparently have not
been studied at all.
Recently, the Whitham modulation equations for
the Kadomtsev-Petviashvili (KP) and two-dimensional Benjamin-Ono
equations and, more generally, a class of (2+1)-dimensional equations
of KP type were derived \cite{ABW2017,ABW2017a,ABR2018}.
The
properties of the resulting KP-Whitham equations were then studied in
\cite{BHM2020} and the soliton limit of these equations was used in
\cite{Ryskamp2021,Ryskamp2021a,Ryskamp2022} to study the time
evolution of a variety of piecewise-constant initial conditions in the
modulation equations and, in the process, characterize the resulting
dynamics of the solutions of the KP equation.
Recently, the Whitham
equations for the radial NLS equation \cite{ACR2019} and those for
focusing and defocusing two-dimensional nonlinear Schr\"odinger
(2DNLS) equations \cite{ACR2021} were also derived using a multiple
scales approach.

The goal of this work is to derive and study the Whitham modulation equations for the defocusing multi-dimensional nonlinear Schr\"odinger equation, which we write in the semiclassical scaling as
\be
i\epsilon \psi_t + \epsilon^2\nabla^2 \psi - 2 |\psi|^2\psi = 0\,
\label{e:nls}
\ee
for a complex-valued field $\psi(\@x,t)$, where
$\@x = (x_1,\dots,x_N)^T$ and
$\nabla^2\psi = \psi_{x_1x_1}+\cdots+\psi_{x_Nx_N}$ is the spatial
Laplacian, and subscripts $x_j$ and $t$ denote partial differentiation
throughout.
Equation~\eqref{e:nls} arises as a governing equation in
water waves \cite{AS1981}, optics \cite{NewellMoloney}, plasmas
\cite{InfeldRowlands}, Bose-Einstein condensates
\cite{KevrekidisFrantzeskakisCarretero2009}, magnetic materials
\cite{ZvezdinPopkov} and beyond.
The small parameter $0<\epsilon\ll1$
quantifies the relative strength of dispersive effects compared to
nonlinear ones and sets a spatial and temporal scale for oscillatory
solutions.
In the (1+1)-dimensional case, the Whitham modulation
equations have been shown to provide quantitative predictions for
experiments in ultracold quantum fluids \cite{PRA74p023623,Hoefer2009}
and nonlinear optics \cite{Wan2007,Xu2017,Bienaime2021,Bendahmane2022}.

While the Whitham equations for the two-dimensional version
of~\eqref{e:nls} (hereafter referred to as the 2DNLS equation) were
obtained in~\cite{ACR2021}, this work differs from \cite{ACR2021} in
several important respects.
First, our derivation employs a two-phase
ansatz for the periodic solutions of the 2DNLS equation, which has
several practical advantages.
For one thing, it immediately yields a
second conservation of waves equation in vector form that was missed
in \cite{ACR2021}.
It is well known that several methods can be used
to derive the Whitham equations: averaged conservation laws, averaged
Lagrangian, and multiple scales perturbation theory.
Our derivation
employs averaged conservation laws which are directly tied to the
physical symmetries of the NLS equation, rather than secularity
conditions as used in \cite{ACR2021}.
Moreover, the ability to take advantage of the second conservation of
waves equation also simplifies the calculations.
In contrast, one of the secularity conditions obtained in \cite{ACR2021}
is equivalent to the averaged energy equation, which is more
complicated and requires more significant manipulation than the second
conservation of waves equation.
Our approach dramatically simplifies the calculations and enables us
to carry out the whole derivation in vector form.
Consequently, the
resulting NLS-Whitham equations are obtained in a simpler way, which
lays the groundwork for generalizations to other NLS-type equations
and higher dimensions.

In this work, we also show how our approach allows one to easily
generalize the derivation of the Whitham equations to the NLS equation
in an arbitrary number of spatial dimensions.
We primarily
concentrate on the two and three dimensional cases, though some of our
results apply to an arbitrary number of spatial dimensions. This
generalization to higher dimensions is particularly relevant because
the NLS equation in three spatial dimensions is the zero-potential
version of the Gross-Pitaevski equation, and is therefore of
fundamental importance in describing the dynamics of Bose-Einstein
condensates \cite{KevrekidisFrantzeskakisCarretero2009}, so we expect
our results to be directly applicable in that context.

We use our representation of the NLS-Whitham equations to identify
several symmetries and reductions of the Whitham equations.
For
example, we verify that the Whitham equations preserve the invariance
of the (N+1)-dimensional NLS equation with respect to scaling and
Galilean transformations, and we take advantage of the vector
formulation of the modulation equations, which we use to show that
they preserve the rotation symmetry of the multidimensional NLS
equation.
We also explicitly write down both the harmonic and soliton limits of
the Whitham equations in a mathematically convenient set of
independent variables (which we refer to as Riemann-type variables)
and in physical variables.
We identify the self-consistent reduction
of the 2DNLS-Whitham equations to the Whitham equations for the radial
NLS equation.

The outline of this work is as follows.
In
section~\ref{s:hydrodynamic} we write the NLS equation in hydrodynamic
form, write down its conservation laws, and obtain a representation
for the periodic solutions.
In section~\ref{s:derivation} we average
the conservation laws to obtain the Whitham equations in physical
variables.
In section~\ref{s:reductions} we begin to study the
reductions of the Whitham equations in physical variables, including
one-dimensional reductions as well as the harmonic and soliton limits.
In section~\ref{s:2DNLSriemann} we discuss two different
transformations to Riemann-type variables.
In
section~\ref{s:furthersymmetries} we derive further symmetries and
reductions of the Whitham equations, including the reduction to the
Whitham equations of the radial NLS equation and
the harmonic and soliton limits of the Whitham equations in Riemann-type variables.
In section~\ref{s:3d} we present the generalization of the results to the NLS equation in three spatial dimensions,
and in section~\ref{s:conclusions} we end this work with a discussion of the results and some final remarks.
The details of various calculations are relegated to the Appendix.

\section{Hydrodynamic form, conservation laws and periodic solutions of the NLS equation}
\label{s:hydrodynamic}

\subsection{Madelung form of the NLS equation and its conservation laws}
We begin by writing down the first few conservation laws of the NLS equation~\eqref{e:nls} in an arbitrary number of dimensions.
It is convenient to introduce the Madelung transformation
\bse
\label{e:madelungansatz}
\bea
\psi(\@x,t) = \sqrt{\rho(\@x,t)}\,e^{i\Phi(\@x,t)}\,,
\\
\@u(\@x,t) = \epsilon\bfnabla \Phi(\@x,t)\,.
\eea
\ese
where 
$\@u = (u_1,\dots,u_N)^T$,
$\@x = (x_1,\dots,x_N)^T$
and $\bfnabla = (\partial_{x_1},\dots,\partial_{x_N})^T$.
Substituting~\eqref{e:madelungansatz} into the NLS equation~\eqref{e:nls}, 
separating into real and imaginary parts, 
and differentiating the real part with respect to each of the spatial variables 
yields the following dispersive hydrodynamic system of PDEs:
\bse
\label{e:hydrodynamicsystem}
\bea
&&\rho_t + 2\bfnabla\cdot(\rho\@u) = 0\,,
\label{e:hydro1}
\\
&&\@u_t + 2(\@u\cdot\bfnabla)\,\@u + 2\bfnabla\rho
  - \txtfrac14 \epsilon^2 \bfnabla\Big(\nabla^2\ln \rho + \frac1\rho \nabla^2\rho\Big) = 0\,.
\label{e:hydro2}
\eea
\ese
The  conservation laws for~\eqref{e:nls} 
for the mass $E$, momentum $\@P$ and energy $H$ in integrated form are:
\bse
\label{e:conservationlaws0}
\be
\deriv{E}t = 0\,,\qquad
\deriv{\@P}t = 0\,,\qquad
\deriv{H}t = 0\,,
\ee
where
\bea E = \int_{\Real^N}|\psi|^2\,(\d\@x)\,,\quad \@P = \txtfrac i2
\epsilon \int_{\Real^N}(\psi\bfnabla \psi^* - \psi^* \bfnabla
\psi)\,(\d\@x)\,,\quad H = \int_{\Real^N}\big( \epsilon^2 \|\bfnabla \psi\|^2 + |\psi|^4 \big)\,(\d\@x)\,,
\nonumber\\
\eea \ese $\|\@v\|^2 = |v_1|^2 + \cdots + |v_N|^2$ is the Euclidean
vector norm and $(\d\@x) = \d x_1\cdots\d x_N$ is the volume element
in $\Real^N$.
These conservation laws correspond, via Noether's
theorem, to the invariance of the NLS equation~\eqref{e:nls} with
respect to phase rotations, space and time translations, respectively
\cite{SulemSulem}.
In differential form, and in terms of the Madelung
variables, these conservation laws become \cite{Jin99}
\bse
\label{e:conservationlaws}
\bea
&&\rho_t + 2\bfnabla\cdot(\rho\@u) = 0\,,
\\
&&(\rho\@u)_t + 2\bfnabla \cdot (\rho\,\@u \otimes \@u) + \bfnabla(\rho^2) = \txtfrac12\epsilon^2\Big(\bfnabla(\nabla^2\rho) - \bfnabla \cdot \Big(\frac{1}{\rho}\nabla\rho \otimes \nabla \rho\Big)\Big)\,,
\\
&&h_t 
  + 2 \bfnabla\cdot\big((h + \rho^2)\@u\big)
  =  \epsilon^2\bfnabla\cdot\Big(\@u\nabla^2\rho - \frac1\rho(\bfnabla\cdot\rho\@u)\nabla\rho \Big)\,,
\eea
\ese
where $\otimes$ denotes the dyadic
[namely, $\@v\otimes\@w = \@v\@w^T$, so that $(\@v\otimes\@w)_{i,j} = v_i w_j$]
and the mass density, momentum density and energy density
of~\eqref{e:nls}
are, respectively 
\vspace*{-1ex}
\bea
\rho = |\psi|^2,\quad 
\rho\@u = i2 \epsilon (\psi\bfnabla \psi^* - \psi^* \bfnabla \psi),\qquad
h = \epsilon^2 \|\bfnabla \psi\|^2 + |\psi|^4 = 
\rho\|\@u\|^2 +\rho^2 + \frac{\epsilon^2}{4\rho} \|\bfnabla\rho\|^2\,.
\nonumber\\[-1ex]
\eea
The first two of the conservation laws~\eqref{e:conservationlaws} 
are equivalent to the real and imaginary parts of NLS equation in hydrodynamic form~\eqref{e:hydrodynamicsystem},
but only up to an extra differentiation, an issue that we will return to later.

\subsection{Periodic solutions of the NLS equation via a two-phase ansatz}
\label{s:periodic}

The Whitham modulation equations govern the slow dynamics of the
parameters of the periodic solutions of the PDE of interest.  
Next, we therefore write down the periodic solutions of the hydrodynamic
system~\eqref{e:hydrodynamicsystem} in arbitrary dimensions.  
We begin by looking for solutions in the form of the following two-phase ansatz:
\bse
\label{e:twophaseanstaz}
\be
\rho(\@x,t) = \rho(Z)\,,
\qquad
\Phi(\@x,t) = \phi(Z) + S\,,
\ee
where $\rho(Z)$ and $\phi(Z)$ are periodic function of $Z$ with period one, and
the ``fast phases'' $Z$ and $S$ are 
\be
Z(\@x,t) = (\@k\cdot\@x - \omega t)/\epsilon\,,\qquad
S(\@x,t) = (\@v\cdot\@x- \mu t)/\epsilon\,.
\ee
\ese
where $\@k = (k_1,\dots,k_N)^T$ and $\@v = (v_1,\dots,v_N)^T$.
The reason for using a two-phase ansatz is the fact that the solution $\psi(\@x,t)$ of the NLS equation~\eqref{e:nls} 
is complex-valued, 
unlike that of the Korteweg-deVries (KdV) equation (of which the KP equation mentioned earlier is a two-dimensional generalization),
which is real-valued.
Therefore, a one-phase ansatz (e.g., as in \cite{ACR2021}) leads only to a subclass of all periodic solutions, 
and one would need to apply a Galilean boost a posteriori in order to capture 
the most general family of periodic solutions of the NLS equation.
Two-phase ansatzes are standard when deriving the Whitham equations using Lagrangian averaging 
(e.g., see \cite{Whitham1974});
the novelty here is that such a two-phase ansatz is combined with the use of averaged conservation laws. 
A key benefit of this approach is the immediate deduction of an additional conservation law 
compared to~\cite{ACR2021}.

In light of~\eqref{e:twophaseanstaz}, the definition \eqref{e:madelungansatz} yields
\be
\label{e:u&v}
\@u(Z) = \@k\,\phi'(Z) + \@v\,,
\ee
using primes to denote derivatives with respect to $Z$ for brevity.
The fact that $\phi(Z)$ is periodic implies 
\vspace*{-0.6ex}
\be
\overline{{\@u}} = \@v\,, 
\ee
where throughout this work the overbar will denote the integral of a
quantity with respect to~$Z$ over the unit period. Moreover, the definition~\eqref{e:madelungansatz}
implies the irrotationality condition
\be
\label{e:uvconstraint}
\bfnabla \wedge \@u = 0\,.
\ee
Hereafter, $\@v\wedge\@w$ is the $N$-dimensional wedge product,
which in two and three spatial dimensions can be replaced by the standard cross product \cite{Bryant,Frankel}.
We substitute the two-phase ansatz \eqref{e:twophaseanstaz} into the hydrodynamic equations~\eqref{e:hydro1} and \eqref{e:hydro2} and collect the leading-order terms, obtaining:
\bse
\label{e:hydrodynamiconephase}
\bea
-\omega \rho'+ 2 \@k \cdot (\rho \@u)'= 0\,, 
\label{e:hleadingh1}
\\ 
-\omega \@u' + 2(\@k\cdot \@u)\,\@u' + 2\@k\rho' - \txtfrac14\Big((\ln \rho)'' + \frac{\rho''}{\rho}\Big)'{\|\@k\|^2}\@k = \@0 \,.
\label{e:hleadingh2}
\eea
\ese
Integrating \eqref{e:hleadingh1} and using \eqref{e:u&v} yields
\be
\phi'(Z) = \frac1{\|\@k\|} \bigg(U +\frac{J}{\rho} - \^{\@k}\cdot\={\@u} \bigg)\,,
\label{e:hphaserelation1}
\ee
where $U = \omega/(2\|\@k\|)$ is the phase speed, $\^{\@k} = \@k/\|\@k\|$, 
and the integration constant~$J$ will be determined later.
Using \eqref{e:hphaserelation1}, we can rewrite~\eqref{e:u&v} as:
\vspace*{-0.4ex}
\be
\@u(Z) = \Big( \frac{J}{\rho} + U \Big)\,\^{\@k} + \={\@u}_\perp\,,
\label{e:uz}
\ee
where 
${\@u}_\perp = \@u - (\^{\@k}\cdot\@u)\,\^{\@k}$.
Importantly, the
requirement that $\phi(Z)$ is periodic implies that $\phi'(Z)$ must
have zero mean.
Taking the inner product of \eqref{e:uz} with $\^\@k$
and averaging the result over the wave period yields a relation
between $\={\@u}$ and $U$, and therefore determines
$\omega = 2\|\@k\| U$: \be
\label{e:v1v2&omega}
U = \^{\@k}\cdot \={\@u} - J\,\overline{\rho^{-1}}\,.
\ee
Next, substituting~\eqref{e:hphaserelation1} into \eqref{e:hleadingh2} and simplifying yields two ODEs for $\rho$.
Note that the two ODEs are consistent thanks to the constraint~\eqref{e:uvconstraint},
which becomes, to leading order, 
\be
\label{e:uvO1}
\@k\wedge\@u' = 0\,.
\ee
Integrating the resulting ODE for $\rho$ one obtains [see Appendix~\ref{a:ODE} for details]
\bea
\label{e:cubicpolynomial}
(\rho')^2 = P_3(\rho)\,,
\label{e:drhodz}
\\
\noalign{\noindent with}
P_3(\rho)  = \frac{4}{\|\@k\|^2} ( \rho - \lambda_1 )( \rho - \lambda_2 )( \rho - \lambda_3 )\,,
\eea
whose solution is
\be
\label{e:rho}
\rho(Z) = A + 4m\|\@k\|^2 K_m^2 \sn^2(2K_m z | m)\,,
\ee
where $A$ is a free parameter, and with
\be
J^2 = A\,\big( A + 4 \|\@k\|^2K_m^2 \big)\big( A + 4 m \|\@k\|^2K_m^2 \big)\,,
\label{e:J2}
\ee
The roots $\lambda_1,\dots,\lambda_3$ are related to the coefficients in the solution~\eqref{e:rho} as
\bea
\lambda_1 = A,\qquad
\lambda_2 = A + 4mK_m^2\|\@k\|^2,\qquad
\lambda_3 =A+4K_m^2 \|\@k\|^2.
\label{e:lambda123_fromAam}
\eea
Conversely, when $\lambda_1,\lambda_2,\lambda_3$ are known, $A$, $\|\@k\|$ and $m$ are given by
\be
A = \lambda_1\,\qquad
\|\@k\|^2 = (\lambda_3 - \lambda_1)/4K_m^2\,,\qquad
m = (\lambda_2 - \lambda_1)/(\lambda_3 - \lambda_1)\,.
\label{e:Aam_from_lambda123}
\ee
The amplitude of the periodic oscillations of the density is
$\lambda_2 - \lambda_1$.  
The requirements $\rho \ge 0$,
$\|\@k\|\ge0$ and $0\le m\le 1$ immediately yield the constraints $A\ge0$ as well as
\be
0 \le \lambda_1\le \lambda_2 \le \lambda_3\,.
\label{e:lambdaconstraint}
\ee
The symmetric polynomials $e_1,\dots,e_3$ defined by the roots $\lambda_1,\dots,\lambda_3$ will also be useful later:
\bea
\label{e:symmetricpolys}
e_1 = \lambda_1 + \lambda_2 + \lambda_3\,,\qquad 
e_2 = \lambda_1 \lambda_2 + \lambda_2\lambda_3 + \lambda_3\lambda_1\,,\qquad 
e_3 = \lambda_1 \lambda_2 \lambda_3 = J^2\,.
\eea
Note that \eqref{e:symmetricpolys} only determines $J$ up to a sign, i.e., $J = \sigma\sqrt{\lambda_1\lambda_2\lambda_3}$,
with $\sigma = \pm1$.
Both sign choices lead to valid solutions of the NLS equation~\eqref{e:nls}.
Some care is deserved when determining the value of $\sigma$ in the
presence of modulations of the periodic solutions, 
as discussed in section~\ref{s:modifiedform}.

The leading-order periodic solution of the hydrodynamic
system~\eqref{e:hydrodynamicsystem} defined
by~\eqref{e:hphaserelation1} and~\eqref{e:rho} contains the following
independent parameters: $A$, $m$, $\@k$, $\={\@u}$ and $\mu$.
However, recall that, to derive the hydrodynamic
equation~\eqref{e:hydro2} from the NLS equation~\eqref{e:nls}, one
differentiates the real part with respect to the spatial variables.
Imposing that the solution of the dispersive hydrodynamic system \eqref{e:hydrodynamicsystem} also solves the
NLS equation (by substituting into the undifferentiated imaginary part
of the NLS equation~\eqref{e:nls})
yields a constraint that determines $\mu$ in terms of the
other constants.
Deriving this relation directly from the above expressions is a bit
cumbersome, but seeking a periodic solution of~\eqref{e:nls} without
writing it in hydrodynamic form [cf.\ Appendix~\ref{s:directperiodic}],
one obtains \be
\label{e:mu}
\mu = 4(1 + m)\|\@k\|^2K_m^2  + 3A + \|\@{\=u}\| - \big(J\overline{\rho^{-1}}\big)^2\,.
\ee
One can now verify that adding this relation to the above solution of
the hydrodynamic system does indeed yield a solution of the NLS
equation~\eqref{e:nls}.
Alternatively, one can obtain~\eqref{e:mu}
using the undifferentiated version of~\eqref{e:hleadingh2}; see
Appendix~\ref{a:ODE}.
Thus, the periodic solutions of the NLS
equation~\eqref{e:nls} in $N$ spatial dimensions contain $2N+2$ scalar
independent parameters: $A$, $m$, $\@k$ and $\@v=\={\@u}$, as one
would expect based on the invariances of the PDE (cf.\
\cite{SulemSulem}).

\subsection{Harmonic and soliton limits of the periodic solutions}

Recall that the harmonic ($m=0$) and soliton ($m=1$) limits of the
Whitham equations for the one-dimensional NLS (1DNLS) equation have
special significance \cite{ElHoefer}.
The same will be true for the
multi-dimensional NLS equation.
It is therefore useful to evaluate
the corresponding limits of the above periodic solutions.

In the limit $m\to0$ (i.e., $\lambda_2\to\lambda_1^+$), 
the solution~\eqref{e:madelungansatz} reduces to a plane wave.  
Indeed, in this limit, we have
\be
\label{e:harmoniccoeffs}
\displaystyle
\rho(Z) = A,\,\qquad
B = 0\,,\qquad
\mu = 2A + \|\={\@u}\|^2\,,\qquad
  J^2=A^2 (\pi^2 \|\@k\|^2 + A )\,,
\ee
and
\be
\psi(\@x,t) = \sqrt{A}\,\,\e^{i(\={\@u}\cdot\@x-(\|\={\@u}\|^2+2A)t)}\,.
\ee
Therefore, the only independent parameters in this case are $A$ and $\={\@u}$. 

In the opposite limit ($m\to1$, i.e., $\lambda_2\to\lambda_3^-$), 
the solution~\eqref{e:madelungansatz} reduces to the soliton solution of the NLS equation.
Indeed, in this limit, \eqref{e:rho} and \eqref{e:Aam_from_lambda123} yield
\bse
\bea
\displaystyle
\rho(Z) = \lambda_1 + (\lambda_3 - \lambda_1) \tanh^2\big[\sqrt{\lambda_3 - \lambda_1}\,\big(\hat{\@k}\cdot \@x - \omega t/\|\@k\|\big)\big],\,\\
B = \lambda_3 - \lambda_1\,,\quad
J^2 = \lambda_1\lambda_3^2\,,\quad
U = \hat{\@k}\cdot\={\@u} - \sigma\sqrt{\lambda_1} \,,\quad
\mu = 2 \lambda_3 + \|\={\@u}\|^2 \,.
\eea
\ese
Note that $\|\@k\|\to0$ as $m\to1$, but $K_m\to\infty$ in such a way that their product remains finite: 
$\|\@k\|K_m \to \sqrt{\lambda_3 -\lambda_1}/2$.
Using \eqref{e:hphaserelation1} we then obtain
\vspace*{-1ex}
\be
\phi(Z) = \arctan\Big[\sqrt{\lambda_3 - \lambda_1} \, \tanh\Big(\sqrt{\lambda_3 - \lambda_1} \,\big(\hat{\@k}\cdot \@x - \omega t/\|\@k\|\big)\Big) / \sqrt{\lambda_1} \big]\,,
\ee
implying 
\vspace*{-1ex}
\be
e^{i \phi + iS } 
  = e^{iS}\Big[\sqrt{\lambda_1} + i\sqrt{\lambda_3-\lambda_1} 
    \tanh \Big(\sqrt{\lambda_3 - \lambda_1} \,\big(\hat{\@k}\cdot \@x - \omega t/\|\@k\|\big)\Big)\Big] \Big/\sqrt{\rho(Z)}\,,
\ee
with $S = \={\@u}\cdot\@x - \mu t$ as before.
Putting everything together, we obtain 
\bea
\label{e:solitonreduction}
\psi(\@x,t) = A_o\e^{-2iA_o^2t}\e^{i(\={\@u}\cdot\@x-\|\={\@u}\|^2t)}
\big\{\cos\theta 
  + i\sin\theta\tanh[A_o\sin\theta\,[\^{\@k}\cdot\@x - 2(\^{\@k}\cdot\={\@u} -A_o\cos\theta)t]]\big\}\,,
\nonumber\\
\eea
with $\={\@u}$ as in \eqref{e:twophaseanstaz},
$A_o = \sqrt{\lambda_3}$ and 
$\theta = \arctan\big[\sqrt{(\lambda_3-\lambda_1)/\lambda_1}\big]$.
The independent parameters of the solution in this case are $\lambda_1$, $\lambda_3$ (or equivalently $A_o$ or $\theta$), 
$\^{\@k}$ 
and $\={\@u}$.
One can further reduce~\eqref{e:solitonreduction} to the more familiar form of the dark soliton solutions of the defocusing NLS equation by choosing $\={\@u} = \@0$.

\section{Derivation of the NLS-Whitham equation by averaged conservation laws}
\label{s:derivation}

We are now ready to study slow modulations of the periodic solutions described above
and derive the Whitham modulation equations that govern them.

\subsection{Nonlinear modulations and averaged conservation laws}

We begin by introducing the following multiple scales ansatz for the solution of the NLS equation~\eqref{e:nls}:
\be
\rho(\@x,t) = \rho(Z,\@X,T)\,,
\qquad
\Phi(\@x,t) = \phi(Z,\@X,T) + S\,,
\label{e:multiplescales}
\ee
where $\@X = \@x $ and $T=t$, with $\rho$ and $\phi$ periodic in $Z$ with period one and 
\bse
\label{e:zSderivs}
\bea
\bfnabla Z = \frac{\@k(\@X,T)}\epsilon\,,\qquad 
Z_t = - \frac{\omega(\@X,T)}\epsilon\,,
\\
\bfnabla S = \frac{\@v(\@X,T)}\epsilon\,,\qquad 
S_t = - \frac{\mu(\@X,T)}\epsilon\,,
\eea
\ese
where, as per the results of section~\ref{s:periodic}, $\@v=\={\@u}$.
The above multiple scales ansatz implies 
\bea
\bfnabla_{\@x} \mapsto \frac {\@k}\epsilon \partial_Z + \frac{\@v}{\epsilon} \partial_S + \nabla_{\@X}\,,\qquad
\partial_t \mapsto - \frac \omega\epsilon \partial_Z - \frac \mu\epsilon \partial_S + \partial_T\,,
\label{e:derivativerules}
\eea
Substituting~\eqref{e:multiplescales} into~\eqref{e:nls}, to leading order we recover the periodic solution~\eqref{e:madelungansatz},
but where all $2N+2$ parameters $A$, $m$, $\@k$ and $\={\@u}$
are now slowly varying functions of $\@X$ and~$t$.
We then seek modulation equations to determine the space-time dependence of these parameters.
To avoid complicating the notation unncessarily,
below we will write derivatives in $\@X$ and $T$ as derivatives in $\@x$ and~$t$.
Equations~\eqref{e:zSderivs} immediately yield the equations of conservation of waves:
\bse
\label{e:averagedconsevationlaws}
\bea
\@k_t + \bfnabla\omega = \@0\,,
\label{e:conservationwaves1}
\\
\bfnabla \wedge \@k = 0\,,
\label{e:curlk}
\\
\={\@u}_t + \bfnabla\mu = \@0\,,
\label{e:conservationwaves2}
\\
\bfnabla \wedge \={\@u} = 0\,.
\label{e:curlu}
\eea
Of course only $N$ equations among~\eqref{e:conservationwaves1} and~\eqref{e:curlk}
are independent,
and similarly for~\eqref{e:conservationwaves2} and~\eqref{e:curlu}.
Equations~\eqref{e:conservationwaves1} and~\eqref{e:conservationwaves2} form the first two vectorial Whitham modulation equations,
whereas~\eqref{e:curlk} and~\eqref{e:curlu} are compatibility constraints.

Next, we obtain the remaining Whitham modulation equations by averaging the conservation laws~\eqref{e:conservationlaws} 
over the fast variable $Z$.
Using~\eqref{e:derivativerules} to replace all spatial and temporal derivatives 
in~\eqref{e:hydrodynamicsystem} and~\eqref{e:conservationlaws},
expanding all terms in powers of $\epsilon$, and averaging, we obtain
at order $\mathcal{O}(\epsilon^0)$
%
\bea
\label{e:avgmass}
(\=\rho)_t + 2 \bfnabla\cdot( \overline{\rho \@u}) = 0\,,
\\
(\overline{\rho \@u})_t + 2 \bfnabla\cdot(\overline{\rho \@u \otimes \@u}) + \bfnabla (\overline{\rho^2})
  + \bfnabla\bigg( \overline{\frac{(\rho')^2}{2\rho}} \,\@k \otimes\@k \bigg) = \@0\,,
\label{e:acl6}
\\
\={h}_t + \bfnabla\cdot\Big( 2\overline{h\@u} + 2\overline{\rho^2\@u}
  + \Big(\@k\cdot\!\overline{\frac{\rho'}{\rho}(\rho\@u)'}\Big)\,\,\@k
  - \|\@k\|^2\overline{\rho''\@u} \Big)
     = 0\,,
\label{e:acl7}
\eea
\ese
where $\=h$ denotes the averaged energy density:
\be
\=h = \overline{\rho\|\@u\|^2}  +\overline{\rho^2} + \frac14{\|\@k\|^2} \overline{{(\rho')^2}/{\rho}}\,.
\label{e:hbar}
\ee
Together with~\eqref{e:conservationwaves1} and~\eqref{e:conservationwaves2},
equations~\eqref{e:avgmass}--\eqref{e:acl7} 
are $3N+2$ scalar PDEs for the $2N+2$ dependent variables $A$, $m$,
$\@k$ and $\@v=\={\@u}$ subject to the $2N$ spatial constraints
\eqref{e:curlk}, \eqref{e:curlu},
and are the desired Whitham modulation equations in physical variables
in any number of spatial dimensions.
Of course, not all of these equations are independent.
We will see later that choosing different subsets of equations still leads to equivalent results,
and in the end the number of independent modulation equations is~$2N+2$.
At the same time, however, we emphasize the simplicity and directness of this approach 
compared to \cite{ACR2021}
in deriving the Whitham equations in multiple spatial dimensions.

\subsection{Modified form of the modulation equations}
\label{s:modifiedform}

In preparation for further simplification of the above system of
Whitham equations, it is convenient to express the periodic solutions
in terms of the roots $\lambda_1$, $\lambda_2$, $\lambda_3$, thereby
replacing $A$, $m$ and $\|\@k\|^2$ as dependent variables.
Explicitly, \eqref{e:uz} and~\eqref{e:rho} become:
\bse
\label{e:periodicparameters}
\bea
\rho(Z) = \lambda_1 + (\lambda_2 -\lambda_1)\,\sn^2(2 K_m z | m)\,,
\\
\label{e:uz2}
\@u(Z) = \@U + \frac{J}{\rho(Z)}\,\^{\@k}\,,
\eea
with 
\bea
\label{e:Udef}
\@U = \={\@u} - J \overline{\rho^{-1}}\,\^{\@k}\,,\qquad
\\
\noalign{\noindent which also implies}
\label{e:omega&mu}
\omega = 2\@k\cdot\@U\,,\qquad
\mu = \lambda_1 +\lambda_2 +\lambda_3 + \|\@U\|^2 + 2U J \overline{\rho^{-1}}\,,
\eea
\ese
with $\^{\@k} = \@k/\|\@k\|$ as before
and $J$, $A$, $\|\@k\|$ and $m$ given in terms of $\lambda_1,\dots,\lambda_3$ by~\eqref{e:J2} and~\eqref{e:Aam_from_lambda123}.
In turn, using~\eqref{e:periodicparameters}, we can write the Whitham modulation equations~\eqref{e:averagedconsevationlaws} as
\bse 
\label{e:WhithamPDE}
\bea
\@k_t + 2\bfnabla (\@k\cdot\@U) = \@0\,,
\label{e:2pw1}
\\
\bfnabla \wedge \@k = 0\,,
\label{e:2pw2}
\\
\big(\@U + J\overline{\rho^{-1}}\,\^{\@k}\big)_t
  + \bfnabla \big(e_1 + \|\@U\|^2 + 2J\overline{\rho^{-1}}\,\@U\cdot\^{\@k} \big) = \@0\,,
\label{e:2pw3}
\\
\bfnabla \wedge  \big( \@U + J\overline{\rho^{-1}}\,\^{\@k} \big) = 0\,,
\label{e:2pw4}
\\
\bar\rho_t + 2 \nabla \cdot \big( J\^{\@k} + \bar\rho \@U \big) = 0\,,
\label{e:2pw5}
\\
(J\,\hat{\@k} + \overline\rho \@U)_t
  + \bfnabla(\overline{\rho^2})
  + \bfnabla\Big[\Big(2\=\rho\@U + 2J\,\hat{\@k}\Big)\otimes \@U 
    + {2J}\@U\otimes\^{\@k}  +  \frac23\big(2e_2 - e_1\=\rho\big)\,\^{\@k} \otimes \^{\@k} \Big] 
  = \@0\,.
\label{e:2pw6a}
\\
\label{e:2pw7}
\bar{h}_t + \bfnabla\cdot\Bigg[
  2J ( 2\, \=\rho + \overline{\|\@u\|^2})\,\^{\@k}  + 2( \overline{\rho^2} + \bar{h} )\,\@U
+ \Bigg( \@U\cdot\^{\@k}\,\Bigg ( \overline{\frac{(\rho')^2}{\rho}}
\Bigg) - \frac{J}{2}
\Bigg( \overline{\frac{(\rho')^2}{\rho^2}} \Bigg)\Bigg)\,\@k \Bigg] 
     = 0\,.
\eea
\ese
See Appendix~\ref{a:ODE} for details on how to obtain~\eqref{e:2pw6a}.
The next step is the evaluation of the elliptic integrals in~\eqref{e:WhithamPDE}.
To this end, we have \cite{NIST}
\vspace*{-1ex}
\bse
\label{e:ellipticintegrals}
\bea
\displaystyle
&&\overline{\rho}=\int_0^1 \rho(Z)\,\d z = \lambda_3 - (\lambda_3-\lambda_1)\frac{\E}{\K}\,,
\label{e:rhobar}
\\
&&\overline{\rho^{-1}} = \int_0^1 \rho^{-1}(Z)\,\d z  = \frac{1}{\lambda_1 K_m}\,\Pi\Big(1-\frac{\lambda_2}{\lambda_1}\Big|m\Big)\,,
\label{e:rho^-1bar}
\eea
\ese
where $\K = K(m)$, $\E=E(m)$ and $\Pi(\cdot|m)$ are the complete
elliptic integrals of the first, second and third kind
respectively.
We also note, for convenience,  that 
\bse
\label{e:avgvelmom}
\bea
\label{e:avgvel}
&&\bar{\@u} = \int_0^1 \@u(Z)\,\d z = \@U + \sigma \frac{\sqrt{\lambda_2\lambda_3}}{\sqrt{\lambda_1}
  \K}\Pi\Big ( 1- \frac{\lambda_2}{\lambda_1} \Big | m \Big ) \^\@k \,
, \\
\label{e:avgmom}
&&\overline{\rho \@u} = \int_0^1 \rho(Z)\@u(Z)\,\d z = \=\rho\@U + J
\^\@k =  \Big
(\lambda_3 - (\lambda_3-\lambda_1) \frac{\E}{\K} \Big )\@U + \sigma \sqrt{\lambda_1\lambda_2\lambda_3} \^\@k\, .
\eea
\ese

We reiterate that not all of the equations~\eqref{e:WhithamPDE} are independent.  
For example, one can obtain~\eqref{e:2pw4} using \eqref{e:2pw3} and~\eqref{e:2pw5}. 
This is relevant because it allows us to work with the most convenient
subset of equations among all the PDEs in~\eqref{e:WhithamPDE},
as long as the compatibility constraints~\eqref{e:2pw2} and~\eqref{e:2pw4} are satisfied.
To this end, recall that $\=h$ is given by~\eqref{e:hbar}, and
\bse
\bea
\overline{\|\@u\|^2} = \|\@U\|^2 + 2J\@U\cdot\^{\@k}\,\overline{\rho^{-1}} + J^2 \overline{\rho^{-2}}\,,
\\
\overline{\rho\|\@u\|^2} = J^2\,\overline{\rho^{-1}} + 2J\@U\cdot\^{\@k} + \=\rho\|\@U\|^2\,.
\eea
\ese
Moreover, the terms $(\rho')^2/\rho$ and $(\rho')^2/\rho^2$, which appear in~\eqref{e:2pw7}, 
can be computed using~\eqref{e:drhodz}. 
On the other hand, 
the averaged energy conservation law~\eqref{e:2pw7} is the most complicated among all of the equations \eqref{e:WhithamPDE}.
In section~\ref{s:3d} we will show that,
thanks to the use of the two-phase ansatz and the resulting second conservation of waves equations~\eqref{e:2pw3} and~\eqref{e:2pw4},
one can avoid having to deal with the averaged energy equation~\eqref{e:2pw7}, 
which greatly simplifies the transformation to Riemann-type variables.

We also point out that the sign of $J$, as
determined by the initial conditions for the system through the value
of $\sigma$---see the discussion
after~\eqref{e:symmetricpolys}---affects $\={\@u}$ via
\eqref{e:avgvel} and $\overline{\rho\@u}$ via \eqref{e:avgmom}.
Therefore, when considering modulations of the periodic solutions, the
value of $\sigma$ depends on $\@x$ and $t$, and its  value must
be chosen in such a way to ensure smoothness of $\overline{\rho\@u}$.
In particular, a sign change of $J$ occurs when the solution hits a
vacuum point, i.e., $\lambda_1=0$.
At such a point, $\=\@u$ is
singular but $\overline{\rho\@u}$ is not.
See \cite{PhysD236p44} for
additional discussion.

\section{Symmetries and reductions of the NLS-Whitham system in physical variables}
\label{s:reductions}

We now present several reductions of the Whitham modulation system~\eqref{e:WhithamPDE} 
in physical variables in arbitrary number of spatial dimensions.
Further symmetries and reductions in the two-dimensional case will be discusssed in section~\ref{s:furthersymmetries}
after we introduce Riemann-type variables in section~\ref{s:2DNLSriemann}.

\subsection{Unidirectional reductions of the modulation equations} 

We begin by showing that the NLS-Whitham equations~\eqref{e:WhithamPDE} reduce to the 1DNLS-Whitham equations
(i.e., the Whitham equations for the 1DNLS equation) when 
$k_2= \cdots = k_N = v_2= \cdots = v_N = 0$
and all quantities are independent of $x_2,\dots,x_N$.
In this case, we have:
\bse
\bea
\|\@k\|^2 = k_1^2\,,\qquad 
u_1(Z) = \frac{J}{\rho} + U\,,\qquad 
\omega =  2k\, U\,,\qquad
U = \=u_1 - J \overline{\rho^{-1}}\,, \qquad
u_2(Z)  = 0\,.
\eea
\ese
The Whitham equations~\eqref{e:2pw2} and \eqref{e:2pw4} and the second
components of~\eqref{e:2pw1}, \eqref{e:2pw3}, and~\eqref{e:2pw5} are satisfied trivially,
while the rest simplify to:
\bse 
\label{e:1DWhithamPDE}
\bea
\label{e:1dw1}
k_t + 2(kU)_x = 0\,,
\\
\label{e:1dw2}
\big( U + J \overline{\rho^{-1}}\big)_t
  + \Big( e_1 + 2J U\,\overline{\rho^{-1}} + U^2 \Big)_x = 0\,,
\\
\label{e:1dw3}
(\=\rho)_t + 2(U\,\=\rho+ J)_x = 0\,,
\\
\label{e:1dw4}
(U\,\=\rho + J)_t
  + \Big(\overline{\rho^2} + 2U^2 \=\rho + 2 J^2 \overline{\rho^{-1}} + \frac{k^2}{2}\overline{ \frac{(\rho')^2}{\rho}}+ 4U\,J\Big)_x = 0\,,
\\
\label{e:1dw5}
\Big(\overline{\rho^2} + U^2 \=\rho + J^2 \overline{\rho^{-1}} + \frac{k^2}{4}\overline{ \frac{(\rho')^2}{\rho}} + 2JU\Big)_t
+ \Big( \frac{3U\,k^2}{2}\overline{\Big( \frac{(\rho')^2}{\rho}\Big)} - \frac{k^2J}{2}\overline{\frac{(\rho')^2}{\rho^2}}
\nonumber
\\
\kern10em
  + 4U\overline{\rho^2} + (4J + 2U^3)\=\rho + 6J U^2 
  + 6J^2U\overline{\rho^{-1}} + 2J^3 \overline{\rho^{-2}}\Big)_x = 0\,,
\eea
\ese
with $x= x_1$.
The system~\eqref{e:1DWhithamPDE} coincides with the modulation equations for the 1DNLS equation~\cite{PRA74p023623}
(cf.\ (4.41) and (4.42) in \cite{PRA74p023623})
upon trivial rescalings resulting from the different normalization 
of the NLS equation in \cite{PRA74p023623} compared to~\eqref{e:nls}.
Note that \eqref{e:1DWhithamPDE} comprise five PDEs for the four solution parameters $A,m,k\&U$ 
(or equivalently $\lambda_1,\lambda_2,\lambda_3,U$).
Once again, one can verify that the modulation equation obtained from~\eqref{e:1dw5} is consistent 
with those obtained from the first four PDEs above.

The above scenario is not the only one in which the Whitham modulation
system~\eqref{e:WhithamPDE} reduces to that of the 1DNLS equation.
Next we consider so-called ``rotated'' one-dimensional reductions
where the rotated coordinate frame is determined by $\_R$, an
$N\times N$ orthogonal matrix.
We introduce the rotated vector
${\@w}^\sharp = \_R\,\@w$ for any vector $\@w$.
Then, the rotated
one-dimensional reduction is obtained through the requirement that $\@k$ and $\={\@u}$ (or equivalently
$\^{\@k}$ and $\@U$) be parallel and that both depend only on $t$ and
the first component of ${\@x}^\sharp$.
We choose $\_R$ so
that $\^{\@k}^\sharp = (1,0,\dots,0)^T$, i.e.,
$k^\sharp_2 = \dots = k^\sharp_N = 0$, which also implies
$U^\sharp_2 = \dots = U^\sharp_N = 0$.
Since the Whitham modulation equations~\eqref{e:WhithamPDE} are
invariant under rotations of the coordinate axes (see below), we
recover the one-dimensional reduction \eqref{e:1DWhithamPDE} when all
quantities are independent of $x^\sharp_2,\dots,x^\sharp_N$ in the
rotated coordinate frame, i.e., with $x$ and all modulation variables in \eqref{e:1DWhithamPDE}
replaced by their rotations $x^\sharp_1$, etc.

\subsection{Invariances of the modulation equations}
\label{s:invariances}

The Whitham modulation equations~\eqref{e:WhithamPDE} are manifestly invariant under 
translations of the spatial and temporal coordinates.
Next we show that the Whitham system~\eqref{e:WhithamPDE} preserves the invariance of the NLS equation
under rotations of the Cartesian coordinates. 
Namely, if $\@x\mapsto\@x^\sharp = \_R \,\@x$, where $\_R$ is an arbitrary constant rotation matrix, 
\eqref{e:WhithamPDE} remain unchanged upon $\@U\mapsto\@U^\sharp = \_R\,\@U$ and $\@k \mapsto \@k^\sharp = \_R\,\@k$.
One can verify that this is indeed the case using the following identities:
\bse
\bea
\displaystyle
\_R\,\bfnabla_{\@x} = \bfnabla_{\@x^\sharp}\,,\quad
 \@U \cdot \@k = \@U^\sharp \cdot \@k^\sharp\,,\quad
 \|\@U\| = \|\@U^\sharp\|\,,\\
 \bfnabla_{\@x}\cdot (\alpha\@k) = \bfnabla_{\@x^\sharp}\cdot (\alpha\@k^\sharp)\,,\quad
 \bfnabla_{\@x}\cdot (\alpha\@U) = \bfnabla_{\@x^\sharp}\cdot (\alpha\@U^\sharp)\,,\\
 \_R\,\bfnabla_{\@x}\cdot (\alpha\@U \otimes \@U) = \bfnabla_{\@x^\sharp}\cdot (\alpha\@U^\sharp \otimes \@U^\sharp)\,,\quad
 \_R\,\bfnabla_{\@x}\cdot (\alpha\@k \otimes \@k) = \bfnabla_{\@x^\sharp}\cdot (\alpha\@k^\sharp \otimes \@k^\sharp)\,,\\
 \_R\,\bfnabla_{\@x}\cdot (\alpha\@k \otimes \@U) = \bfnabla_{\@x^\sharp}\cdot (\alpha\@k^\sharp \otimes \@U^\sharp)\,.
\eea
where $\alpha$ is an arbitrary real number.
\ese

Next we show that the Whitham system~\eqref{e:WhithamPDE} also preserves the invariance of the NLS equation
with respect to scaling and spatial reflections and Galilean transformations.
Recall that, if $q(\@x,t)$ is any solution of the NLS equation, so are
$q^\sharp(\@x,t) = \alpha q(\alpha\,\@x,\alpha^2 t)$, 
$q^\sharp(\@x,t) = q( -\@x, t)$ and
$q^\sharp(\@x,t) = q(\@x - 2\@w t , t)e^{i (\@w \cdot \@x - \|\@w\|^2 t)}$ where all transformation parameters are real-valued.
We next show that the modulation equations \eqref{e:WhithamPDE} are invariant under each of these transformations. 
Specifically, letting $q^\sharp(\@x,t) = [\rho^\sharp(\@x,t)]^{1/2}\,e^{i \phi^\sharp(\@x,t)}$, we have,
for the scaling symmetry,
\be
\label{e:scalingperiodicsolution}
\rho^\sharp(\@x,t) = \alpha^2 \rho(\alpha\@x,\alpha^2t)\,,
\qquad
\phi^\sharp(\@x,t) =  \phi(\alpha\@x,\alpha^2t)\,,
\ee
and the dependent variables of the Whitham equations become
\bse
\label{e:scalingvariable}
\bea
\lambda^\sharp_j(\@x,t) = \alpha^2 \lambda_j(\alpha \@x, \alpha^2 t)\,,\, j = 1,2,3\,, 
\\
\@k^\sharp(\@x,t) = \alpha \@k(\alpha \@x, \alpha^2 t)\,,
\quad
\@U^\sharp(\@x,t) = \alpha \@U(\alpha \@x, \alpha^2 t)\,,
\quad
J^\sharp(\@x,t) = \alpha^3 J(\alpha \@x, \alpha^2 t)\,.
\eea
\ese 
Using \eqref{e:scalingperiodicsolution} and \eqref{e:scalingvariable}, one can show that the Whitham modulation equations~\eqref{e:WhithamPDE} remain unchanged.
Similarly, it can be seen that spatial reflections leave the modulation equations invariant 
upon the following transformation of the dependent variables:   
\bse
\label{e:spartialvariable}
\bea
\rho^\sharp(\@x,t) = \rho(-\@x,t)\,,
\quad
\lambda^\sharp_j(\@x,t) = \lambda_j(-\@x,t)\,,\, j = 1,2,3\,, \quad
\@k^\sharp(\@x,t) = -\@k(-\@x,t)\,,
\quad\\
\@U^\sharp(\@x,t) = -\@U(-\@x, t)\,,
\quad
J^\sharp(\@x,t) = J(-\@x, t)\,.
\eea
\ese 
Finally, with regards to Galilean transformations, 
writing $q^\sharp(\@x,t) = \sqrt{\rho^\sharp(\@x,t)}\,e^{i\phi^\sharp(\@x,t)}$ implies
\be
\rho^\sharp(\@x,t) = \rho(\@x - 2 \@w t, t)\,,
\qquad
\phi^\sharp(\@x,t) =  \phi(\@x - 2 \@w t, t) + \@w \cdot \@x - \|\@w\|^2 t\,.
\ee
The dependent variables in the modulation equations \eqref{e:WhithamPDE} become
\bse
\label{e:galileanvariable}
\bea
\lambda^\sharp_j(\@x,t) = \lambda_j(\@x - 2 \@w t, t)\,,\, j = 1,2,3\,, 
\\
\@k^\sharp(\@x,t) = \@k(\@x - 2 \@w t, t)\,,
\quad
\@U^\sharp(\@x,t) = \@U(\@x - 2 \@w t, t) + \@w\,,
\quad
J^\sharp(\@x,t) = J(\@x - 2 \@w t, t)\,.
\eea
\ese 
Using \eqref{e:galileanvariable}, one can verify that the modulation equations~\eqref{e:WhithamPDE} remain invariant under the above Galilean transformation. 
The Riemann-type variables, which will be introduced in section~\ref{s:2DNLSriemann}, change as follows under the above transformations:
\bea
r_j^\sharp(\@x ,t) =\alpha r_j(\alpha \@x, \alpha^2 t),\quad
r_j^\sharp(\@x ,t) = r_j(-\@x,t),\quad
r^\sharp_j(\@x ,t) = r_j(\@x - 2 \@w t, t) + \@w \cdot \hat{\@k}/2,\quad j = 1,2,3,4\,.
\nonumber\\
\eea

\subsection{Harmonic and soliton limits of the modulation equations in physical variables}

The harmonic and soliton limits of the Whitham equations for the KdV
and 1DNLS equations have proven to be quite useful to study various
nonlinear dynamical scenarios of practical
interest~\cite{Congy2019,PRL120p144101,Sprenger2018}.
The same is
true for the harmonic and soliton limits of the KP-Whitham
equations~\cite{BHM2020,Ryskamp2021,Ryskamp2021a,Ryskamp2022}.
We
therefore expect that the same will also be true for the harmonic and
soliton limit of the NLS equation in multiple spatial dimensions.

Like with the periodic solution, the harmonic limit of the Whitham
equations is the limit $m\to0$, corresponding to
$\lambda_2\to\lambda_1^+$.
Recall that in this limit the solution
becomes a plane wave.
The integrals in~\eqref{e:ellipticintegrals} simplify considerably:
\vspace*{-1ex}
\be
\overline{\rho} = \lambda_1\,,\quad \overline{\rho^2} = \lambda_1^2\,,\quad \Big(\overline{\frac{(\rho')^2}{\rho}}\Big) =0\,,\quad\overline{\rho^{-1}} = 1/\lambda_1\,,\quad \overline{\rho^{-2}} = 1/\lambda_1^2\,,
\qquad
J = \sigma \lambda_1 \sqrt{\lambda_3} \,.
\ee
Then, the linear dispersion relation is
\be
\label{e:omega_m=0}
\omega = 2 \|\@k\| \left (\^{\@k}\cdot\={\@u} - \sigma \sqrt{\pi^2
    \|\@k\|^2 + \=\rho^2}\, \right )\,,
\ee
the averaged energy limits
to $\=h = \=\rho \|\=\@u\|^2 + \=\rho^2$ and the Whitham
equations~\eqref{e:WhithamPDE} reduce to:
\bse
\label{e:W_m=0}
\bea
\@k_t + \bfnabla \omega = \@0\,,
\label{e:w1_m=0}
\\
\={\@u}_t + \bfnabla \big( 2 \=\rho + \|\={\@u}\|^2 \big) = \@0\,,
\qquad
\bfnabla \times \={\@u} = 0\,,
\label{e:w4_m=0}
\\
\=\rho_t + 2\bfnabla\cdot (\=\rho\={\@u} ) = 0\,,
\label{e:w5_m=0}
\\
(\=\rho\={\@u})_t + \bfnabla(\=\rho^2)  + 2\bfnabla \cdot \big( \=\rho \={\@u} \otimes \={\@u}\big) = \@0\,,
\label{e:w6_m=0}
\\
\label{e:w7_m=1}
\=h_t + \nabla \cdot \big ( 2 (\=h + \=\rho^2) \=\@u \big ) = 0\, .
\eea
\ese
Again, not all of these equations are independent.  
For example, one can derive~\eqref{e:w6_m=0} 
using~\eqref{e:w4_m=0} and \eqref{e:w5_m=0}.
Also, note that the variable $\@k$ is immaterial, since its value does not affect the solution,
and~\eqref{e:W_m=0} is decoupled from the other PDEs.
Thus, equations~\eqref{e:w4_m=0} and~\eqref{e:w5_m=0},
which are equivalent to the shallow water equations, 
are by themselves a closed system of evolution PDEs 
for the parameters of the plane wave solution, $\=\rho$ and $\={\@u}$.
Nonetheless, \eqref{e:w1_m=0} describes the evolution of a harmonic wave propagating on top of the mean flow.

Finally, we discuss the soliton limit of the Whitham modulation system~\eqref{e:WhithamPDE},
obtained for $m\to1$ corresponding to $\lambda_2 \to \lambda_3$.
In this limit, 
the integrals in~\eqref{e:ellipticintegrals} become:
\be
\overline{\rho} = \lambda_3\,,\quad 
\overline{\rho^2} = \lambda_3^2\,,\quad 
\|\@k\|^2\,\overline{\Big(\frac{(\rho')^2}{\rho}\Big)} =0\,,\quad
\overline{\rho^{-1}} = 1/\lambda_3\,,\quad 
\overline{\rho^{-2}} = 1/\lambda_3^2\,.
\ee
Then~\eqref{e:2pw1} and~\eqref{e:2pw2} are trivially satisfied, and the rest simplify to:
\bse 
\label{e:vectorformU&Vm=1}
\bea
\displaystyle
\label{e:vw1_m=1}
&&
\={\@u}_t + \nabla \big(2\=\rho + \|\={\@u}\|^2 \big) = \@0\,,
\\
\label{e:vw2_m=1}
&&
\=\rho_t + 2 \nabla \cdot ( \=\rho \={\@u} ) = 0\,,
\\
\label{e:vw3_m=1}
&&
(\=\rho\={\@u})_t + \nabla(\=\rho^2)  + 2\nabla \cdot \big( \={\@u}
\otimes \=\rho\={\@u}\big) = \@0\,,
\\
\label{e:vw4_m=1}
&&
\=h_t + \nabla \cdot \big ( 2 (\=h + \=\rho^2) \=\@u \big ) = 0\, .
\eea
\ese
Note that, similar to before, we can derive equation~\eqref{e:vw3_m=1}
and \eqref{e:vw4_m=1} from \eqref{e:vw1_m=1} and \eqref{e:vw2_m=1}.
Therefore, we have a system of $N+2$ PDEs for the dependent variables
$\={\@u}$ and $\=\rho=\lambda_2$.
But in this case, we are missing
PDEs for $\lambda_1$ and $\^{\@k}$ that define the soliton amplitude
and its propagation direction, which are needed to completely
determine the soliton solution.
This deficiency is also present in
the one-dimensional case.
The one-dimensional case is simpler,
however, because, in that case, $\@k$ is a one-component vector, and
therefore $\^{\@k} = \pm 1$, constant.
The soliton limit is singular
so care must be taken in its calculation.
In any case, both in the
one-dimensional and higher-dimensional situation, the problem is
eliminated by the transformation to Riemann-type variables, as we will
see later.

\section{2DNLS-Whitham equations in Riemann-type variables}
\label{s:2DNLSriemann}

In this section and the next one we 
temporarily restrict our attention to the two-dimensional case and perform suitable changes of dependent variables 
to simplify the form of the 2DNLS-Whitham equations.

When $N=2$, the modulation system \eqref{e:WhithamPDE} consists of
eight PDEs for six dependent variables in the independent variables
$\@x = (x,y)^T$ and $t$, plus the two scalar
constraints~\eqref{e:2pw2} and~\eqref{e:2pw4}.
We will use the four
scalar conservation of waves equations~\eqref{e:2pw1} and
\eqref{e:2pw3} together with the averaged conservation of mass
\eqref{e:2pw5} and one of the components of the conservation of
momentum equations~\eqref{e:2pw6a}, neglecting the compatibility
conditions~\eqref{e:2pw2} \& \eqref{e:2pw4} as well as the
conservation of energy~\eqref{e:2pw7}.
Importantly, however, the
resulting Whitham equations are equivalent to those obtained by
working with a different set of averaged equations \cite{ACR2021}.

As in the one-dimensional case, the transformation involves two steps.  
The first step is the change of dependent variables
from $(A,k_1,k_2,m,\=u_1,\=u_2)$ to
$\@Y = (\sqrt{\lambda_1},\sqrt{\lambda_2},\sqrt{\lambda_3},U_1,U_2,q)$, 
with
\be
q = k_2/k_1 = \tan\varphi\,,
\label{e:qdef}
\ee
similar to~\cite{ABW2017},
where $\varphi = \arctan(k_2/k_1)$
[not to be confused with the fast phase $\phi(Z)$ that was used in sections~\ref{s:hydrodynamic} and~\ref{s:derivation}]
identifies the direction of the periodic wave's fronts:
\be
\^{\@k} = (\cos\varphi,\sin\varphi)^T\,.
\label{e:khat2D}
\ee
The second step of the transformation is then defined by the map from $\lambda_1,\lambda_2,\lambda_3$ and $U_1$ to 
the ``Riemann-type'' variables $\.r_1,\.r_2,\.r_3,\.r_4$ via the transformation%
\vspace*{-0.2ex}
\bse
\bea
U_1 = \txtfrac12 \cos \varphi\,(\.r_1 + \.r_2 + \.r_3 + \.r_4),
\\
\lambda_1 = \txtfrac14 (\.r_1 - \.r_2 - \.r_3 + \.r_4)^2,\quad
\lambda_2 = \txtfrac14 (\.r_1 - \.r_2 + \.r_3 - \.r_4)^2,\quad
\lambda_3 = \txtfrac14 (\.r_1 + \.r_2 - \.r_3 - \.r_4)^2\,.
\label{e:VfromR}
\eea
\ese
The variables $\.r_1,\dots,\.r_4$ are one possible two-dimensional generalization of the Riemann invariants of
the Whitham equations for the 1DNLS equation.
Note that in this work the overdot does not denote differentiation with respect to time.

Recall that the existence of Riemann invariants for (1+1)-dimensional hydrodynamic-type systems 
is intimately tied to the integrability properties of the modulation equations.
Using the one-dimensional Riemann invariants as dependent variables in higher-dimensional systems
diagonalizes their one-dimensional reductions, 
and makes the equations more advantageous for analysis (e.g., see \cite{ElHoefer}).
We will show below that,
for both the two-dimensional and three-dimensional cases, 
a suitable generalization of the one-dimensional Riemann invariants allows one to 
write the modulation equations in a concise and convenient form.

In terms of $\.r_1,\dots,\.r_4$, the periodic solution~\eqref{e:rho} becomes%
\bse
\label{e:m&rhoinr}
\bea
\rho(Z)= \txtfrac14 (\.r_1-\.r_2 -\.r_3 +\.r_4)^2 + (\.r_2-\.r_1)(\.r_4-\.r_3)\sn^2(2\K\,Z|m)\,,\\
m =\frac{(\.r_2-\.r_1)(\.r_4-\.r_3)}{(\.r_3-\.r_1)(\.r_4-\.r_2)}\,.
\eea
\ese
Moreover, $\.{\@R} = (\.r_1,\.r_2,\.r_3,\.r_4,U_\perp,q)^T$ satisfies the hydrodynamic system
\be
\label{e:riemannwhitham}
\.{\@R}_t + \_M_1\.{\@R}_x + \_M_2\.{\@R}_y = 0\,.
\ee
The matrices $\_M_1$ and $\_M_2$ are rather complicated, and we therefore omit them for brevity.
When $k_2 = U_\perp = 0$, however,
the last two equations in~\eqref{e:riemannwhitham} are trivially satisfied, and the first four 
reduce to the Whitham equations for the 1DNLS equation in Riemann invariant (diagonal) form \cite{forestlee,pavlov}:
\be
\partialderiv{\.{\@r}}t + \_V\,\partialderiv{\.{\@r}}x = 0\,,
\label{e:WhithamRiemann1DNLS}
\ee
with 
\be
\.{\@r} = (\.r_1,\dots,\.r_4)^T,\qquad 
\.{\_V} = \diag(\.{\@V}),\qquad
\.{\@V} = (\.V_1,\dots,\.V_4)^T,\qquad
\ee
\bse
\label{e:1Driemannspeeds}
\bea
\.V_1 = 2V_o + \frac{2(\.r_2-\.r_1)(\.r_4-\.r_1)\K}{(\.r_4-\.r_2)\E - (\.r_4-\.r_1)\K}\,,\quad
\.V_2 = 2V_o + \frac{2(\.r_2-\.r_1)(\.r_3-\.r_2)\K}{(\.r_3-\.r_2)\K - (\.r_3-\.r_1)\E}\,,
\\[0.4ex]
\.V_3 = 2V_o + \frac{2(\.r_3-\.r_2)(\.r_4-\.r_3)\K}{(\.r_4-\.r_2)\E - (\.r_3-\.r_2)\K}\,,\quad
\.V_4 = 2V_o + \frac{2(\.r_4-\.r_1)(\.r_4-\.r_3)\K}{(\.r_4-\.r_1)\K - (\.r_3-\.r_1)\E}\,,
\eea
\ese
\par\medskip\noindent
with $V_o = U_1$.

The Whitham modulation system~\eqref{e:riemannwhitham} can be further simplified 
by introducing a modified set of Riemann-type variables:
\bse
\be
r_j = \cos\varphi\,\,\.r_j\,,\quad j=1,\dots,4\,,
\label{e:rtildefromrdot}
\ee
with $q = \tan\varphi$ as before.
Moreover, the curl-free constraint~\eqref{e:curlu} yields [see section~\ref{s:3d} for details]
\be
p = \sec\varphi\,\,U_\perp\,,
\label{e:rtransf}
\ee
\ese
where the perpendicular component of $\@U$ is defined by
\be
\label{e:Uperp}
U_\perp = \@U\cdot \^\@k_\perp\,, \qquad \^\@k_\perp = (-\sin
\varphi,\cos\varphi)^T\, .
\ee
The Whitham modulation equations~\eqref{e:riemannwhitham}
then reduce to the following form:
\be
\partialderiv{\@R}t + \_A~\partialderiv{\@R}x + \_B~\partialderiv{\@R}y = 0\,,
\label{e:Whitham}
\ee
where 
$\@R = (r_1,\dots,r_4,q,p)^T$,
\bse
\label{e:Whithamcoeffs}
\be
\_A = \left( 
\begin{array}{cc}
A_{4\times4} & A_{4\times2}\\
A_{2\times4} & A_{2 \times 2}
\end{array}\right)\,,
\qquad
\_B =\left( 
\begin{array}{cc}
B_{4\times4} & B_{4 \times 2}\\
B_{2 \times 4} & B_{2 \times 2}
\end{array}\right)\,,
\ee
with $g = 1+q^2$ as in \cite{ACR2021}
and 
\bea
A_{4\times4} = \_V - q^2 U_1 \,\Ones_4 + q^2 (\@1\otimes\@r + \@r\otimes\@1)\,,
\quad
A_{2\times2} = 2
\left( 
\begin{array}{cc} (1-q^2)U_1 & -q^2  \\ q^2 (2U_1^2 - s_2) & gU_1 \end{array}\right)\,,
\\
A_{4\times2} = -q\big(~ 2U_1\@r - \_V\@r + \@a/g ~,~ U_1\@1 - (2(q^2-1)\@r + \@V) /g ~\big)\,,~~
A_{2\times4} = gq\,\big(\, -\@1 ~,~ 2(U_1\@1 - \@r) ~\big)^T,
\nonumber\\
\\
B_{4\times4} = q(\_V + U_1 \,\Ones_4) + 2p\,\_I_4 - q (\@1\otimes\@r + \@r\otimes\@1)\,,
\quad
B_{2\times2} = 2\left( \begin{array}{cc} p + 2q\,U_1 & q  \\ -q (2U_1^2 - s_2) & p \end{array}\right)\,,
\\
B_{4\times2} = 1/(2g) \,\big(\, 2\@a ~,~ (1-q^2)(4\@r - \@V) \,\big)\,,\qquad
B_{2\times4} = - A_{2\times4}/q\,,
\eea
with 
\bea
\@r = (r_1,\dots,r_4)^T,\\
\_V = \diag(\@V),\quad
\@V = (V_1,\dots,V_4)^T,\quad
\label{e:Vtildedef}
\\
\@a = \txtfrac13 \big[ 4U_1(1-3q^2)\@r - 2U_1\@V - (1+3q^2)((s_2 - 2U_1^2)\@1 - \_V\@r) \,\big]\,,
\label{e:alpha_j}
\eea
\ese
where $U_1 = (r_1 + r_2 + r_3 + r_4)/2$,
$V_1,\dots,V_4$ are as in \eqref{e:1Driemannspeeds}
but with $(r_1,\dots,r_4)$ instead of $(\.r_1,\dots,\.r_4)$,
$\@1 = (1,\dots,1)^T$,
$\_I_n$ is the $n\times n$ identity matrix,
$\Ones_n$~denotes the $n \times n$ matrix with all entries equal to one, 
and
\be
s_n = r_1^n + r_2^n + r_3^n + r_4^n\,.
\label{e:sdef}
\ee
In component form, the Whitham modulation equations~\eqref{e:Whitham} are \cite{ACR2021}
\bse
\label{e:Whitham2Ddiagonal}
\bea
\partialderiv{r_j}t + V_j \partialderiv{r_j}x + (qV_j+2p)\partialderiv{r_j}y + h_j = 0,   \quad j=1,2,3,4, \label{e:Whithamdiagonal1to4}
\\
\partialderiv qt + 2 \big(g\,U_1 + pq \big) \partialderiv qx + 2 \Dy{} \big[ g U_1 + pq \,\big] = 0\,,
\label{e:diagonalq}
\\
\partialderiv pt + 2 g\,U_1 \partialderiv px + 2p \partialderiv py + \Dy{} \big[ g(s_2 -2U_1^2) \,\big] = 0\,,
\label{e:diagonalp}
\eea
\ese
where 
\bse
\bea
h_j = 2q(U_1-r_j)\Dy{U_1} - \frac12q\Dy{s_2} + q(V_j-2U_1)\bigg( r_j \partialderiv qx + \frac12\partialderiv px \bigg)
  + \frac{a_j}{g} \Dy q  - \frac{1-q^2}{2g}(V_j-4r_j) \Dy p\,
\nonumber\\
\label{e:hsimple}
\eea
and $D_y$ is the ``convective'' derivative as in \cite{ABW2017}:
\be
\Dy{} = \partialderiv{}y - q\partialderiv{}x\,.
\label{e:Dy}
\ee
\ese
The steps to obtain~\eqref{e:Whitham2Ddiagonal} are just a special case of the ones needed to simplify the Whitham equations
for the three-dimensional NLS equation, which will be discussed in section~\ref{s:3d}.
All the calculations in section~\ref{s:3d} can be trivially reduced to the two-dimentional case
by simply taking $(q_1,q_2) = (q,0)$ and $(p_1,p_2) = (p,0)$ there. 
Therefore we omit the details here for brevity.

Note the necessary compatibility condition for equations~\eqref{e:Whitham2Ddiagonal} 
in which the initial data is subject to the curl-free constraints $\nabla\times\=\@u = \nabla\times\@k = 0$,
similarly to the KP equation \cite{ABW2017,ACR2021}.
In section~\ref{s:3d} we will show how these constraints can be written out explicitly in terms of the Riemann-type variables.

\section{Further symmetries and reductions of the 2DNLS-Whitham equations}
\label{s:furthersymmetries}

Both of the sets of Riemann-type variables $\.\@R$ and $\@R$ introduced in section~\ref{s:2DNLSriemann}
are useful to study further symmetries of the 2DNLS-Whitham system.

\subsection{Reduction to the Whitham equations for the radial NLS equation}

The Whitham equations for the 2DNLS equation admit a self-consistent reduction to the Whitham equations 
for the radial NLS equation, which were recently derived~\cite{ACR2019}.
To show this, we first perform a change of independent variables from the Cartesian coordinates $x$ and $y$ to 
the polar coordinates 
\vspace*{-1ex}
\be
\label{e:polar}
R = \sqrt{x^2+y^2}\,,
\qquad
\theta = \arctan(y/x) \,.
\ee
Using the definition of the convective derivative $D_y$
in~\eqref{e:Dy}, we find
\be
\Dy f = \frac{(y-qx)}{R}\,\partialderiv{f}{R}
  + \frac{(x+qy)}{R^2} \partialderiv{f}{\theta}\,.
\ee
Equations~\eqref{e:diagonalq} and~\eqref{e:diagonalp} in polar coordinates then become, respectively,
\bse
\label{e:qpequations_polar}
\bea
\fl
q_t + g \sum_{i=1}^4 \big[(\sin\theta - q\cos\theta) (r_i)_R + \frac{(\cos\theta + q \sin\theta)}{R} (r_i)_{\theta}\big]
  + 2q (\sin\theta - q\cos\theta) p_R 
  + \frac{2q}{R} (\cos\theta + q \sin\theta)p_{\theta} 
\nonumber\\
\fl\kern4em
  + \big[2U_1 (1-q^2) \cos\theta
  + 2 (p + 2qU_1)\sin\theta\big]q_R
  + \big[2(p+2qU_1)\cos\theta -2U_1(1-q^2)\sin\theta\big]\frac{q_{\theta}}{R} = 0\,,
\label{e:dqdt_polar}
\\
\fl
p_t + 2g \sum_{i=1}^4 (r_i-U_1)\big[(\sin\theta - q\cos\theta)(r_i)_r + (\cos\theta + q\sin\theta )\frac{(r_i)_{\theta}}{r}\big]
  + 2(gU_1\cos\theta + p\sin\theta)p_R 
\nonumber\\
\fl\kern4em
  + 2(p\cos\theta - gU_1\sin\theta)\frac{p_{\theta}}{R}
  + 2q(s_2 -2U_1^2)\big[(\sin\theta - q\cos\theta)q_R + (\cos\theta + q\sin\theta )\frac{q_{\theta}}{R}\big] = 0\,.
\label{e:dpdt_polar}
\eea
\ese
We then look for a reduction of~\eqref{e:qpequations_polar} and the remaining four Whitham equations~\eqref{e:Whithamdiagonal1to4} 
in which $q =\tan\theta = y/x$. 
With this assumption, 
\eqref{e:qpequations_polar} simplify considerably.
We~also seek solutions in which the Riemann-type variables 
$\.r_1,\dots,\.r_4$ are independent of the angular coordinate $\theta$. 
Recall that the variables $r_1,\dots,r_4$
appearing in \eqref{e:qpequations_polar} 
are related to $\.r_1,\dots,\.r_4$ by~\eqref{e:rtildefromrdot}.
Thus
\be
\label{e:derivativeoftildes}
    \partialderiv{r_i}R = \frac1{\sqrt{g}}\partialderiv{\.r_i}R -\frac{qr_i}{g^{3/2}}\partialderiv qR\,,
    \qquad
    \partialderiv{r_i}{\theta} = -\frac{q\.r_i}{g^{3/2}}\partialderiv q\theta\,.
\ee
Substituting the above expression into~\eqref{e:dqdt_polar} 
and~\eqref{e:dpdt_polar} yields, respectively,
\bse
\bea
p_{\theta} + \cot\theta p =0\,,
\label{e:radialconstraint3}
\\
p_t + 2(U_1 \sec \theta + p\sin\theta) p_R + 2(p\cos\theta -\tan\theta \sec\theta U_1)\,{p_{\theta}}/{R} = 0\,.
\label{e:radialconstraint4}
\eea
\ese
Equation~\eqref{e:radialconstraint3} yields $p(R,\theta,t) = C(R,t)\,\csc\theta$,
with $C(R,t)$ to be determined.
Then, substituting this expression into~\eqref{e:radialconstraint4} yields 
$C_t + 2(U_1 \sec \theta + C)\,C_R - 2(C\,\cot^2\theta - U_1 \sec \theta)\,{C}/{R} = 0$,
whose only self-consistent solution is $C=0$, implying $p(R,\theta,t) = 0$.

Now we turn our attention to the reduction of the first four Whitham modulation equations, namely~\eqref{e:Whithamdiagonal1to4}. 
Tedious but straightforward calculations show that, when written in the polar coordinates~\eqref{e:polar}, 
and using $q = \tan\theta$ and $p=0$  
as well as~\eqref{e:derivativeoftildes},
the four modulation equations~\eqref{e:Whithamdiagonal1to4} become
exactly the Whitham equations for the radial NLS equation derived in~\cite{ACR2019}:
\vspace*{1ex}
\be
\partialderiv{\.{\@r}}t + \.{\_V}\,\partialderiv{\.{\@r}}R + \frac{\.{\@b}}{R} = 0\,,
\label{e:WhithamRadialNLS}
\ee
\vspace*{1ex}
with $\.{\@r} = (\.r_1,\dots,\.r_4)^T$ and $\.{\_V} = \diag(\.{\@V})$ as in section~\ref{s:2DNLSriemann}, 
with $\.{\@b} = (\.b_1,\dots,\.b_4)^T$, 
\bse
\label{e:betadef}
\bea
\displaystyle
\.b_1 &= 2V_o^2 - \txtfrac13{(\.r_2 + \.r_3 + \.r_4)V_1}
    -\txtfrac13{[(\.r_2 + \.r_3)^2 + (\.r_3 + \.r_4)^2 + (\.r_2 + \.r_4)^2]}\,,
\\
\.b_2 &= 2V_o^2 - \txtfrac13{(\.r_1 + \.r_3 + \.r_4)V_2} 
    -\txtfrac13{[(\.r_1 + \.r_3)^2 + (\.r_3 + \.r_4)^2 + (\.r_1 + \.r_4)^2]}\,,
\\
\.b_3 &= 2V_o^2 - \txtfrac13{(\.r_1 + \.r_2 + \.r_4)V_3} 
    -\txtfrac13{[(\.r_1 + \.r_2)^2 + (\.r_2 + \.r_4)^2 + (\.r_1 + \.r_4)^2]}\,,
\\
\.b_4 &= 2V_o^2 - \txtfrac13{(\.r_1 + \.r_2 + \.r_3)V_4} 
    -\txtfrac13{[(\.r_1 + \.r_2)^2 + (\.r_2 + \.r_3)^2 + (\.r_3 + \.r_1)^2]}\,,
\eea
\ese
and $V_o = \half(\.r_1+\.r_2+\.r_3+\.r_4)$ as before.
In terms of the physical variables, the assumption $q = \tan\theta$ implies that the wavefronts 
are oriented radially, 
and the requirement $p=0$ means that the mean flow has no transversal component either,
which are both conditions that are consistent with a radially symmetric reduction.

\subsection{Harmonic limit and soliton limit of the 2DNLS-Whitham equations in Riemann-type variables}
\label{s:harmonicsolitonlimits}

In section~\ref{s:reductions} we studied the harmonic limit and the soliton limit of the modulation equations 
in physical variables, and we saw that the singular soliton limit yields fewer equations than are needed
to describe the parameters of the soliton solutions of the 2DNLS equation.
We next study the corresponding limits of the Whitham modulation equations in Riemann-type variables, 
and we show how the transformation to Riemann-type variables eliminates this problem and yields a closed system of equations.


The harmonic limit ($m\to0$) corresponds to either $r_2\to r_1^+$ or $r_3\to r_4^-$.
In the former case, the PDE~\eqref{e:Whithamdiagonal1to4} with $j=1$ and the one with $j=2$
coincide, as needed for the limit to be a self-consistent reduction, 
and the Whitham modulation system~\eqref{e:Whitham} then becomes 
\bse
\be
\@R_t + \_A_{o.1}\@R_x + \_B_{o.1}\@R_y = 0\,,
\ee
with $\@R = (r_1,r_3,r_4, q,p)^T$.
The matrices $\_A_{o.1}$ and $\_B_{o.1}$ are simply the matrices 
$\_A$ and $\_B$
from section~\ref{s:2DNLSriemann} with $r_2=r_1$ and the second row and column omitted.  
Moreover, the Riemann speeds reduce to
\vspace*{-0.4ex}
\be
    V_1 = V_2 = 4r_1 -\frac{(r_3 - r_4)^2}{2r_1 -r_3 -r_4}\,,\qquad 
    V_3 = 3r_3  + r_4\,,\qquad
    V_4 = r_3 + 3r_4\,,
\ee
while $h_1,\dots,h_4$ are still given by~\eqref{e:hsimple} with $r_2 = r_1$.
In the latter case (i.e., $r_3\to r_4^-$), the PDE~\eqref{e:Whithamdiagonal1to4} with $j=3$ and the one with $j=4$
coincide, 
and the Whitham modulation system~\eqref{e:Whitham} then becomes 
\bse
\be
\@R_t + \_A_{o.2}\@R_x + \_B_{o.2}\@R_y = 0\,,
\ee
with $\@R = (r_1,r_2,r_3, q,p)^T$.
The matrices 
$\_A_{o.2}$ and $\_B_{o.2}$ are just the matrices $\_A$ and $\_B$
from section~\ref{s:2DNLSriemann} with $r_4=r_3$ and the fourth row and column omitted.  
The Riemann speeds reduce to
\vspace*{-0.4ex}
\be
    V_1 = 3r_1 + r_2\,,\qquad 
    V_2 = r_1 + 3r_2\,,\qquad
    V_4 = V_3 = 4r_3 + \frac{(r_1-r_2)^2}{r_1 + r_2 -2r_3} \,,
\ee
with $h_1,\dots,h_4$ now given by~\eqref{e:hsimple} with $r_4 = r_3$.
In both cases, it is straightforward to verify that, 
once the transformation to Riemann-type variables is inverted and the modulation equations 
are written back in terms of the physical variables, one recovers the system~\eqref{e:W_m=0}.


The soliton limit ($m\to1$) corresponds to $r_3\to r_2^+$.
In this case, the PDEs~\eqref{e:Whithamdiagonal1to4} with $j=3$ and the one with $j=2$ coincide, 
and the remaining equations become
\bse
\be
\@R_t + \_A_1\@R_x + \_B_1\@R_y = 0\,,
\ee
with $\@R = (r_1,r_2,r_4, q,p)^T$.
The matrices $\_A_1$ and $\_B_1$ are $\_A$ and $\_B$
from section~\ref{s:2DNLSriemann} with $r_3=r_2$ and the fourth row and column
omitted. 
The Riemann speeds reduce to
\be
    V_1 = 3r_1 + r_4\,,\qquad 
    V_2 = V_3 = r_1 + 2r_2 + r_4\,\qquad
    V_4 =  r_1 + 3r_4\,,
\ee
\ese
where $h_1,\dots,h_4$ are still given by~\eqref{e:hsimple} with $r_3 = r_2$.
As in the harmonic limit, it is straightforward to verify that, 
once the transformation to Riemann-type variables is inverted and the modulation equations 
are written back in terms of the physical variables, one recovers the system~\eqref{e:vectorformU&Vm=1}.
In this case, however, the equations in Riemann-type variables also allow us to obtain 
the two previously missing modulation equations,
which determine the evolution of $\^{\@k}$ and the soliton amplitude $a = \lambda_3 - \lambda_1$.
One of these equations is immediate, since \eqref{e:diagonalq} directly determines $q = \tan\varphi$ 
and therefore $\^{\@k}$.
As for the amplitude equation, note that \eqref{e:VfromR} yields
$\lambda_3 - \lambda_1 = \sec^2\varphi\,(r_4-r_2)(r_3-r_1)$.
Therefore, the modulation equation for $r_1$, $r_2 = r_3$, $r_4$,
and $q$ determine the evolution of the soliton amplitude and direction.

\section{Whitham modulation equations for the NLS equation in three spatial dimensions}
\label{s:3d}

We now show how, thanks to the rotation-invariant form of all equations in sections~\ref{s:hydrodynamic} 
and~\ref{s:derivation},
the results of section~\ref{s:2DNLSriemann} are easily generalized to the NLS equation in three spatial dimensions.

\subsection{Set-up and resulting 3DNLS-Whitham system}

The Madelung transformation~\eqref{e:madelungansatz}
yields the same hydrodynamic system of PDEs~\eqref{e:hydrodynamicsystem} as well as
the mass, momentum and energy conservation laws~\eqref{e:conservationlaws}
in differential form,
now with $\@u = (u_1,u_2,u_3)^T$, $\@x = (x,y,z)^T$
and $\bfnabla = (\partial_x,\partial_y,\partial_z)^T$.
The two-phase ansatz~\eqref{e:twophaseanstaz} is also the same, 
now with 
$\@k = (k_1,k_2,k_3)^T$ and $\@v = (v_1,v_2,v_3)^T$,
and the curl-free condition~\eqref{e:uvconstraint} is now $\nabla\times\@u = 0$.
The only difference is the number of independent parameters in the periodic solutions:
eight in three spatial dimensions as opposed to six in two spatial dimensions.
The whole derivation in section~\ref{s:derivation} also remains the same,  
including the averaged conservation laws~\eqref{e:averagedconsevationlaws} 
and the Whitham modulation equations~\eqref{e:WhithamPDE},
again the only difference being the number of equations, which in
three dimensions is eleven evolutionary equations.

The first point at which the derivation for the three-dimensional case diverges from the two-dimensional one
is the transformation to Riemann-type variables. 
Compared to~\cite{ACR2021}, the process here is made much easier 
by the availability of the second conservation of waves equation \eqref{e:conservationwaves2}, 
which allows us to bypass the averaged conservation of energy,
which, in turn, greatly simplifies the calculations even in the presence of a third spatial dimension.
We begin with the natural generalization of the parametrization~\eqref{e:khat2D} for $\^{\@k}$, namely:
\bse
\bea
\^{\@k} = (\cos\varphi,\sin\varphi\cos\alpha,\sin\varphi\sin\alpha)^T\,.
\label{e:khat3D}
\\
q_1 = k_2/k_1 = \tan\varphi\,\cos\alpha\,,\quad 
q_2 = k_3/k_1 = \tan\varphi\,\sin\alpha\,,\quad
\label{e:qdef3D}
\\
g = 1 + q_1^2 + q_2^2 = 1/k_1^2 = \sec^2\varphi\,.
\label{e:gdef3D}
\eea
\ese
The leading-order part~\eqref{e:uvO1} of the curl-free
condition~\eqref{e:uvconstraint} now consists of three equations.
The first two of them are
$k_1u_2' = k_2 u_1'$ and $k_1 u_3' = k_3 u_1'$, 
which, when integrated, yield
\be
\@u_\flat(Z) = (u_2(Z),u_3(Z))^T = u_1(Z)\,\@q + \@p\,,
\label{e:u2&u3}
\ee
with 
$\@q = (q_1,q_2)^T$, $\@p = (p_1,p_2)^T$,
and $p_1$, $p_2$ are additional modulation variables depending on the slow
variables $\@x$ and $t$ that appear due to integration in $Z$.
For any three-component vector~$\@w = (w_1,w_2,w_3)^T$
we introduce the ``flat'' notation $\@w_\flat = (w_2,w_3)^T$, 
which we use extensively,
 to denote the two-component vector comprised of the
second and third components of the vector $\@w$.
The third equation, $k_2u_3' = k_3 u_2'$ is automatically satisfied.
Also, averaging~\eqref{e:u2&u3}, we obtain the two additional relations:
\bse
\label{e:ubar3D}
\bea
\={\@u}_\flat = \=u_1\@q + \@p\,,
\\
\@U_\flat = U_1\@q + \@p\,.
\eea
Similarly, the first component of~\eqref{e:uz2} 
yields 
\bea
u_1(Z) = U_1 + J k_1/(\|\@k\|\rho(Z))\,,
\label{e:u1&U1}
\\
\noalign{\noindent
together with}
\omega = 2k_1 (g U_1 + \@q\cdot\@p)\,.
\label{e:omega3D}
\eea
\ese
Finally, we define the Riemann-type variables $r_1,\dots,r_4$ via the same transformation as in section~\ref{s:2DNLSriemann},
namely:
\vspace*{-0.4ex}
\bse
\label{e:lambda123_to_r1234}
\bea
U_1 = \txtfrac12(r_1 + r_2 + r_3 + r_4),
\\
\lambda_1 = \txtfrac14\,g(r_1 - r_2 - r_3 + r_4)^2,\quad
\lambda_2 = \txtfrac14\,g(r_1 - r_2 + r_3 - r_4)^2,\quad
\lambda_3 = \txtfrac14\,g(r_1 + r_2 - r_3 - r_4)^2.
\nonumber\\
\eea
\ese
Then, in sections~\ref{s:derivation3d}, \ref{s:convective3d} and~\ref{s:riemann3d} below, we show that the 
Whitham modulation equations~\eqref{e:WhithamPDE} yield the eight-component system of equations
\vspace*{0.6ex}
\bse
\label{e:Whitham3D}
\bea
\partialderiv{\@r}t + \_V\,\partialderiv{\@r}x + ( \@q\otimes\_V +2\@p\otimes\_I_4) \cdot \bfnabla_\flat \@r
  + \@h(\@r,\@q,\@p) = 0,   
\label{e:3D:diagonalW1to4}
\\
\partialderiv{\@q}t + 2(U_1 + \@q\cdot\@U_\flat)\,\partialderiv{\@q}x + 2\@D_\flat\,( U_1 + \@q\cdot\@U_\flat ) = 0\,,
\label{e:dqdt3D}
\\
\partialderiv{\@p}t + 2(U_1 + \@q\cdot\@U_\flat)\partialderiv{\@p}x + \@D_\flat\,( g(\~e_1 - U_1^2) + \|\@p\|^2 ) = 0\,.
\label{e:dpdt3D}
\eea
\ese
Here, as before, $\@r = (r_1,\dots,r_4)^T$, $\_V = \diag(\@V)$ with $\@V = (V_1,\dots,V_4)^T$ as in~\eqref{e:Vtildedef},
and the dot product in~\eqref{e:3D:diagonalW1to4} operates on the two-component vectors to its left and its right.
That is, in component form, for each $j=1,\dots,4$
the third term in~\eqref{e:3D:diagonalW1to4} is the dot product between $\@q\,V_j +2\@p$ and $\bfnabla_\flat r_j$.
Additionally, 
\eqref{e:dqdt3D} and~\eqref{e:dpdt3D} contain the three-dimensional 
generalization of the convective derivative of \cite{ABW2017} and section~\ref{s:2DNLSriemann}, namely:
\bse
\be
\label{e:Dperpdef}
\@D_\flat = (D_y,D_z)^T = \bfnabla_\flat - \@q\,\partial_x\,,
\ee
where $\bfnabla_\flat = (\partial_y,\partial_z)^T$
and 
\bea
\Dy{} = \partialderiv{}y - q_1\partialderiv{}x\,,\qquad  
\Dz{} = \partialderiv{}z - q_2\partialderiv{}x\,.
\label{e:Dy&Dz}
\eea
\ese
The term $\@h(\@r,\@q,\@p) = (h_1,\dots,h_4)^T$ in~\eqref{e:3D:diagonalW1to4}
is given by 
\bse
\bea
h_j = 2 (U_1-r_j) \@q\cdot\@D_\flat U_1 - \txtfrac12 \@q\cdot\@D_\flat s_2
  + (V_j -2U_1) \,\@q\cdot \Big(r_j \partialderiv{\@q}{x} + \frac12 \partialderiv{\@p}{x}\Big)
  - \txtfrac14( V_j  - 4r_j) \,\@D_\flat \cdot \@p 
\nonumber
\\\kern2em
+ a_j \@D_\flat \cdot \@q 
  + (b_j/g)\, \tr[(\@q\otimes\@q)(\@D_\flat\otimes\@q)]
  + ((V_j-4r_j)/g)\, \tr[(\@q\otimes\@q)(\@D_\flat\otimes\@p)],
\label{e:3D:h_j}
\\[1ex]
\noalign{\noindent with}
a_j = \txtfrac13 [2(2r_j-V_j)U_1 - s_2 + 2U_1^2 + V_j r_j],
\quad
b_j = r_j (V_j-4U_1) - s_2 +2U_1^2 + a_j,
\eea
\ese

\smallskip
\noindent 
for $j=1,\dots,4$.
The $s_n$ are as in~\eqref{e:sdef},
and $\~e_1 = g(\lambda_1 + \lambda_2 + \lambda_3)$ in
\eqref{e:dpdt3D}, is similar to~\eqref{e:symmetricpolys}. 
Equations~\eqref{e:3D:diagonalW1to4}, \eqref{e:dqdt3D}, \eqref{e:dpdt3D} and~\eqref{e:3D:h_j} should be 
compared to~\eqref{e:Whithamdiagonal1to4}, \eqref{e:diagonalq}, \eqref{e:diagonalp} and~\eqref{e:hsimple} 
in the two-dimensional case.
Note that, while $\@h_4(\@r,\@q,\@p)$ might give the impression of a forcing term in~\eqref{e:3D:diagonalW1to4}, 
that is not the case in reality, 
as \eqref{e:3D:h_j} shows that 
$\@h_4(\@r,\@q,\@p)$ is in fact a homogenous first-order differential polynomial in~$\@r$, $\@q$ and $\@p$,
and therefore \eqref{e:Whitham3D} is indeed a system of PDEs of hydrodynamic type
like its one-dimensional and two-dimensional counterparts.

Similarly to the two-dimensional case, the 
3DNLS-Whitham modulation equations \eqref{e:Whitham3D} are subject
to the compatibility conditions
$\nabla\times\=\@u(\@x,0) = \@0$ and $\nabla\times\@k(\@x,0) = \@0$
at $t = 0$. 
In Appendix~\ref{a:3Dsimplification} we show that, in terms of the dependent variables defined above, these constraints 
become, respectively,
\be
k_1\@q_x = {\@D_\flat k_1}\,,\qquad
k_1\@p_x = 2 ((\bfnabla_{\@r}{k_1})^T\_R_4)\cdot\@D_\flat\@r - 2U_1 {\@D_\flat k_1}\,,
\label{e:curlfreeRiemann}
\ee
where 
$\bfnabla_{\@r} = (\partial_{r_1},\dots,\partial_{r_4})^T$,
$\_R_4 = \diag(r_1,\dots,r_4)$,
and the dot product operates on the four-component vectors to its left and its right. 
Equations~\eqref{e:curlfreeRiemann} are conditions that must be satisfied by 
the initial conditions for~\eqref{e:Whitham3D}
in order for its solutions to represent modulations of actual one-phase solutions of the NLS equation~\eqref{e:nls}.

\subsection{Derivation of the 3DNLS-Whitham system: Equations for the auxiliary variables}
\label{s:derivation3d}

To derive the evolution equation~\eqref{e:dqdt3D} for $\@q$, we split the first conservation of waves equation~\eqref{e:conservationwaves1} and rewrite it
using the convective derivatives $D_y$ and $D_z$ defined in~\eqref{e:Dy&Dz}, to obtain
\be
\label{e:conservationwaves1p&q}
k_{1,t} + \omega_x = 0\,,
\qquad
{\@q}_t + (\@D_\flat\omega)/k_1 = 0\,,
\qquad
{\@q}_x = (\@D_\flat k_1)/k_1\,,
\ee
with $\omega$ as in~\eqref{e:omega3D}.
The first of equations~\eqref{e:conservationwaves1p&q} will be used later to derive~\eqref{e:3D:diagonalW1to4}.
Substituting the third equation in~\eqref{e:conservationwaves1p&q}
into the second one and using~\eqref{e:omega3D}
yields the desired evolution equation~\eqref{e:dqdt3D}.
Note also that the third equation in~\eqref{e:conservationwaves1p&q} is precisely the first of the 
constraints~\eqref{e:curlfreeRiemann}.

Next, to derive the evolution equation~\eqref{e:dpdt3D} for $\@p$, we start with the constraint~\eqref{e:curlu}
for the second conservation of waves equation. 
Using~\eqref{e:ubar3D}, \eqref{e:curlu} yields
\be
(\=u_1\@q + \@p)_x = \bfnabla_\flat \=u_1\,.
\ee
Averaging~\eqref{e:u1&U1} over the unit period, we can rewrite the above as 
\be
\partial_x \big[( U_1 + J\,\overline{\rho^{-1}}/g^{1/2})\@q + \@p \big] = \bfnabla_\flat \big(U_1 + J\,\overline{\rho^{-1}}/g^{1/2}\big)\,,
\ee
and simplifying further we obtain
\be
{\@p}_x = \@D_\flat U_1 + \@D_\flat\big(J\,\overline{\rho^{-1}}/g^{1/2}\big) - \big(U_1 + J\,\overline{\rho^{-1}}/g^{1/2}\big)(\@D_\flat k_1)/{k_1}\,.
\label{e:p_x}
\ee
Now we use the second conservation of waves~\eqref{e:conservationwaves2}, written in the form of~\eqref{e:2pw3}.
From the second and third components, together with the above relations, we have
\bea
\partial_t \big( \big(U_1 + J\,\overline{\rho^{-1}}/g^{1/2}\big)\@q + \@p \big) + 
  \bfnabla_\flat \big( g(U_1^2 + \~e_1) + \|\@p\|^2 + 2U_1\@q\cdot\@p + 2 (gU_1 + \@q\cdot\@p) J\,\overline{\rho^{-1}}/g^{1/2}\big) = 0\,.
\nonumber\\
\eea
Simplifying yields
\bea
{\@p}_t + \@D_\flat\big(g(U_1^2 + \~e_1) + \|\@p\|^2 + 2U_1\@p \cdot \@q\big) - 2U_1\, \@D_\flat (gU_1 + \@p \cdot \@q) 
\nonumber
\\
\kern10em
+ 2 (gU_1 + \@p \cdot \@q) \big( \@D_\flat\big(J\,\overline{\rho^{-1}}/g^{1/2}\big) - \big(U_1 + J\,\overline{\rho^{-1}}/g^{1/2}\big)(\@D_\flat k_1)/{k_1}  \big) =0\,.
\nonumber\\
\eea
Using~\eqref{e:p_x} yields the desired equation~\eqref{e:dpdt3D}.

\subsection{Derivation of the 3DNLS-Whitham system: Convective derivatives}
\label{s:convective3d}

It remains to derive the four equations in~\eqref{e:3D:diagonalW1to4}
for the Riemann-type variables $r_1,\dots,r_4$.
To this end, we use the two conservation of waves equations \eqref{e:2pw1} and~\eqref{e:2pw3} 
(as well as the compatibility conditions~\eqref{e:2pw2} and~\eqref{e:2pw4}) 
along with the averaged conservation of mass and momentum equations \eqref{e:2pw5}, \eqref{e:2pw6a}. 
The process comprises three main steps.

The first step is the further simplification of the averaged conservation laws.
Note that 
\bea
k_1^2 = g(\lambda_3-\lambda_1)/4\K^2\,.
\label{e:C0}
\eea
For convenience we also introduce the quantity
$\@M = \overline{\rho\@u}/g = (M_1,M_2, M_3)^T$, 
with
\be
M_1 := \overline{\rho u_1}/g = U_1 \overline{\rho}/g + \~J\,,
\qquad
\@M_\flat := M_1 \@q + (\overline\rho/g)\, \@p = (\overline\rho/g) \@U_\flat + \~J \@q\,, 
\label{e:M}
\ee
and $\~J = \^k_1\,J/g$.
Then, using~\eqref{e:omega3D} one can rewrite the modulation equations~\eqref{e:2pw1} and \eqref{e:2pw5} as follows:
\bea
k_{1,t} + 2 [k_1 (U_1 + \@q \cdot \@U_\flat)] = 0\,,
\label{e:k1_t}
\\
(\overline\rho)_t + 2(gM_1)_x + 2 \bfnabla_\flat \cdot (g~\@M_\flat) = 0\,,
\label{e:rhobar_t}
\eea
while the first component of the second conservation of waves equation~\eqref{e:2pw3} becomes
\be
(U_1 + J\,\overline{\rho^{-1}}/g^{1/2})_t + \big[ g\big(\~e_1 + U_1^2 + 2J\,\overline{\rho^{-1}}/g^{1/2} U_1\big) + \|\@p\|^2 + 2(U_1 + J\,\overline{\rho^{-1}}/g^{1/2}) \@p \cdot \@q\big]_x = 0\,.
\nonumber
\label{e:(u+s)_t}
\ee
Moreover, using the equation~\eqref{e:barrho^2} we can write the averaged momentum equation~\eqref{e:2pw6a} in component form as
\bse
\bea
(g M_1)_t + \big[g\big(2U_1 (M_1 + \~J) + \~e_2 - \overline\rho^2/g^2 \big) + \overline\rho^2 \big]_x 
\nonumber\\\kern10em
  + \bfnabla_\flat \cdot \big[ 2g (M_1 + \~J)\@U_\flat - 2g\~J \@p + g(\~e_2 - \overline\rho^2/g^2) \@q  \big] =0\,,
\label{e:M1_t}
\\
(g \@M_\flat)_t + \Big[g \big( 2(M_1 + \~J)  \@U_\flat + (\~e_2 - \overline\rho^2/g^2) \@q - 2\~J \@p \big)\Big]_x + \bfnabla_\flat\overline\rho^2) 
\nonumber\\\kern10em
  + \bfnabla_\flat \cdot \Big[ 2g \@M_\flat \otimes \@U_\flat + 2g\~J \@U_\flat \otimes \@q + g(\~e_2 - \overline\rho^2/g^2) \,\@q \otimes \@q  \Big] = \@0\,.
\label{e:M2&M3}
\eea
\ese

Next, we perform a second, intermediate step to write the Whitham modulation equations in terms of convective derivatives.
First, we derive some identities that will be useful later.
Equation~\eqref{e:2pw2} and the definition of $\@q$ in~\eqref{e:qdef3D} yield
\be
\@q_x = \frac{1}{k_1} \@D_\flat k_1 \,,
\quad
\@q_y = \frac{1}{k_1} \@D_\flat (k_1q_1)\,,
\quad
\@q_z = \frac{1}{k_1} \@D_\flat (k_1q_2)\,.
\label{e:qidentity}
\ee
Moreover, in Appendix~\ref{a:3Dsimplification}, we show that
these relations also yield the two constraints
\bse
\bea
D_y q_2 = D_z q_1\,,
\label{e:qrelation}
\\
D_y p_2 = D_z p_1\,,
\label{e:prelation}
\eea
\ese
which will prove to be useful.
We then define the additional convective derivatives
\be
\label{e:Dx}
D_x  = \partialderiv{}x +\@q \cdot \bfnabla_\flat\,,
\qquad
D_t  = \partialderiv{}t + 2U_1\partialderiv{}x + 2 \@U_\flat \cdot \bfnabla_\flat\,.
\ee

Now we rewrite the evolution equations for $\@q$ using these convective derivatives. 
Specifically, in Appendix~\ref{a:3Dsimplification} we show that
\eqref{e:k1_t} and~\eqref{e:(u+s)_t} yield, respectively,
\bea
\label{e:DK1}
\frac{D_t k_1}{k_1} + 2 D_x U_1 + W_1 = 0\,,
\\
\label{e:D(U+S)}
D_t(U_1+J\,\overline{\rho^{-1}}/g^{1/2}) + (U_1-J\,\overline{\rho^{-1}}/g^{1/2}) \frac{D_tk_1}{k_1} + D_x(\~e_1 + U_1^2) + 2W_2 = 0\,,
\eea
where
\bse
\bea
gW_1 = \@q \cdot [D_t \@q + 2U_1\,D_x\@q + 2D_x\@p] \,,    
\label{e:gW_1}
\\
gW_2 = \@q \cdot \big[U_1\,D_t \@q + s_2 D_x \@q + \frac12 D_t \@p + U_1\,D_x \@p\big] \,.
\label{e:gW_2}
\eea
\ese
Moreover, in Appendix~\ref{a:3Dsimplification}
we also show that the conservation of mass equation \eqref{e:rhobar_t}
and conservation of momentum equation~\eqref{e:M1_t} yield, respectively,
\bea
g\Big[D_t (\overline\rho/g) - (\overline\rho/g) \frac{D_t k_1}{k_1} + 2D_x \~J \Big] + (\overline\rho/g) (\@q \cdot D_t \@q) + 6 \~J \@q \cdot D_x \@q 
\nonumber\\\kern8em
  + 2M_1 \big(g \bfnabla_\flat \cdot \@q - \@q \cdot D_x \@q \big) + 2 (\overline\rho/g) \big(g \bfnabla_\flat \cdot \@p - \@q \cdot D_x \@p \big) =0\,,
\label{e:W3}
\\
g \Big[ (\overline\rho/g) D_t U_1 + D_t \~J - 2\~J \frac{D_t k_1}{k_1} + D_x \~e_2\Big] + \@q \cdot [M_1 D_t \@q + (\overline\rho/g) D_t \@p + 4 \~e_2 D_x \@q ]
\nonumber\\\kern8em
+ (\~e_2 - (\overline\rho^2/g^2) + 2U_1 \~J) \big( g \bfnabla_\flat \cdot \@q - \@q \cdot D_x \@q\big) + 2\~J \big( g \bfnabla_\flat \cdot \@p - \@q \cdot D_x \@p \big) =0\,.
\nonumber\\
\label{e:W4}
\eea
Equations~\eqref{e:DK1}, \eqref{e:D(U+S)}, \eqref{e:W3} and \eqref{e:W4} comprise the four modified modulation equations written in terms of the variables $U_1$, $\lambda_1$, $\lambda_2$ and $\lambda_3$ and the convective derivatives $D_x$ and~$D_t$.

\subsection{Derivation of the 3DNLS-Whitham system: Equations for Riemann-type variables}
\label{s:riemann3d}

The third and final step in the derivation of~\eqref{e:3D:diagonalW1to4}
is to express the modulation equations in terms of $r_1,\dots,r_4$.
Recall the transformation~\eqref{e:lambda123_to_r1234} to the Riemann-type variables.
Note that the arrangement of indices in~\eqref{e:lambda123_to_r1234} is dictated by the requirement that the constraint~\eqref{e:lambdaconstraint}
be satisfied when $r_1\le r_2\le r_3\le r_4$, since
\bse
\bea
\lambda_2 - \lambda_1 = g(r_4-r_3)(r_2-r_1)\,,~~
\\
\lambda_3 - \lambda_1 = g(r_4-r_2)(r_3-r_1)\,,~~
\\
\lambda_3 - \lambda_2 = g(r_4-r_1)(r_3-r_2)\,.
\eea
\ese
In Appendix~\ref{a:3Dsimplification}, 
using the above definitions,
we show that
\eqref{e:DK1}, \eqref{e:D(U+S)}, \eqref{e:W3} and \eqref{e:W4} yield,
respectively, 
\bse
\label{e:waveW1234}
\bea
\label{e:waveW1}
(\bfnabla_{\@r}[\,\log{k_1}])^T D_t \@r + D_xs_1 + W_1 &=& 0\,,
\\
\label{e:waveW2}
2(\bfnabla_{\@r}[\,\log{k_1}])^T \_R_4 \,D_t\@r + D_xs_2 + 2 W_2 &=& 0\,,
\\
\label{e:massW3}
3(\bfnabla_{\@r}[\,\log{k_1}])^T \_R_4^2 D_t\@r + D_xs_3 + 3W_3 &=& 0\,,
\\
\label{e:momentumW4}
4(\bfnabla_{\@r}[\,\log{k_1}])^T \_R_4^3 \,D_t\@r + D_xs_4 + 4W_4 &=& 0\,,
\eea
\ese
where $\@r = (r_1,\dots,r_4)^T$,
$\bfnabla_{\@r} = (\partial_{r_1},\dots,\partial_{r_4})^T$
and 
$\_R_4 = \diag(r_1,\dots,r_4)$ as before, 
with $W_1$ and $W_2$ as in~\eqref{e:gW_1} and~\eqref{e:gW_2}, 
and 
\bse
\bea
gW_3 = \txtfrac14(s_2-2U_1^2) gW_1 + U_1 gW_2 + \txtfrac12 \@q \cdot [(\overline\rho/g) D_t \@q + 6\~J D_x \@q] 
\nonumber\\\kern4em
  + (U_1\=\rho/g +\~J ) \big(g (\bfnabla_\flat \cdot \@q) - \@q \cdot D_x \@q \big) 
  + (\overline\rho/g) \big(g (\bfnabla_\flat \cdot \@p) - \@q \cdot D_x \@p \big)\,,
\label{e:gW3}
\\
gW_4 = \txtfrac18 (6\~J - U_1 s_2 + 2U_1^3 ) gW_1 + \txtfrac14 (s_2 - 4 U_1^2)gW_2 +  \txtfrac32 U_1 gW_3 
\nonumber\\\kern4em
 + \txtfrac14 \big[ \@q \cdot (M_1 D_t \@q + (\overline\rho/g) D_t \@p + 4\~e_2 D_x \@q)
\nonumber\\\kern4em
 + (\~e_2-(\overline\rho^2/g^2) + 2U_1\,\~J)\big(g (\bfnabla_\flat \cdot \@q) - \@q \cdot D_x \@q \big) 
 + 2\~J \big(g (\bfnabla_\flat \cdot \@p) - \@q \cdot D_x \@p \big)  \big]\,.
\label{e:gW4}
\eea
\ese
Importantly, note that, even though the second conservation of waves equation \eqref{e:D(U+S)}
contains the third complete elliptic integral $\Pi(\,\cdot\,,m)$ via $\overline{\rho^{-1}}$
[cf.\ \eqref{e:rho^-1bar}],
the third elliptic integral does not appear in the resulting
modulation equation~\eqref{e:waveW1}.
Note that
$\Pi(\cdot,m)$ is also contained in the conservation of energy equation.
Next, one can collect the four equations~\eqref{e:waveW1234} and rewrite them in matrix form as
\be
\label{e:Whithammatrix}
\_M(\@r)\,\big( \nabla_{\@r}[\log k_1]\cdot D_t\@r + D_x\@r \big) + \@W= \@0\,,
\ee
where 
$\@W=(W_1, \cdots, W_4)^T$ 
and $\_M(\@r)$ is the Vandermonde matrix
\be
\_M(\@r) = \left(
\begin{array}{cccc}
   1 &1&1&1 \\
   r_1 & r_2 & r_3 & r_4 
   \vspace{1mm}
   \\
   r_1^2 & r_2^2 & r_3^2 & r_4^2
   \vspace{1mm}
   \\
   r_1^3 & r_2^3 & r_3^3 & r_4^3 
\end{array} \right)\,.
\ee
Multiplying \eqref{e:Whithammatrix} by $\_M^{-1}(\@r)$,
we then finally obtain~\eqref{e:3D:diagonalW1to4}, with
\bea
h_j = \frac{(-1)^{j+1}\Delta_{ilm}}{|\Delta|\,(\partial k/\partial r_j)/k}[r_ir_lr_mW_1 - (r_ir_l+r_lr_m+r_mr_i)W_2 + (r_i+r_l+r_m)W_3 - W_4],~~ 
j=1,\dots,4\,,    
\nonumber\\
\label{e:3D:hj_Cramer}
\eea
where $j\neq i, j\neq l, j\neq m$, $i<l<m$, summation of repeated indices is implied, and
\be
|\Delta| = \prod_{j>l}^4(r_j-r_l)\,,  
\quad
\Delta_{ilm} = (r_i-r_l)(r_l-r_m)(r_m-r_i)\,.
\ee
Finally, using equations \eqref{e:Dq} and \eqref{e:Dp}, one can simplify $h_1,\dots,h_4$ in~\eqref{e:3D:hj_Cramer} to obtain~\eqref{e:3D:h_j}.

\section{Discussion and perspectives}
\label{s:conclusions}

In summary, we derived the Whitham modulation equations for the defocusing NLS equation in two, three and higher
spatial dimensions using a two-phase ansatz and the averaged conservation laws of the NLS equation
written in coordinate-free vector form,
and we elucidated various symmetries and reductions of the resulting equations, including
the reduction to the Whitham equations of the radial NLS equation as well as 
the harmonic and soliton limits. 
We point out that, long after this work was completed,
we learned that modulation equations for multi-dimensional equations of NLS type were written down 
in physical variables using a general framework in \cite{Rodrigues2022},
and the modulation equations were used to study the stability of the plane wave solutions.
On the other hand, no transformation to Riemann-type variables was carried out in \cite{Rodrigues2022}.

We reiterate that the use of a two-phase ansatz in this work (as opposed to a one-phase ansatz as in \cite{ACR2021})
greatly simplifies the derivation, since it results in a second conservation of waves equation that 
allows us to avoid using the conservation of energy equation, 
which is much more complicated in comparison.
Moreover, the advantage of using a two-phase ansatz increases with the number of spatial dimensions.
This is because the number of modulation equations needed is $2N+2$.
Therefore, if one tried to derive the modulation equations in three spatial dimensions with a one-phase ansatz,
one would need to use additional conservation laws for the NLS equation.
This would not only lead to a much more complicated derivation, 
but one would quickly exhaust the number of available conservation laws,
since the NLS equation in more than one spatial dimensions is not completely integrable, 
and therefore does not have hidden symmetries resulting in an infinite number of conservation laws.

In contrast, the results of section~\ref{s:3d} can be generalized in a
straightforward way to obtain the Whitham modulation equations 
in simplified form in an arbitrary number of spatial dimensions.
The system of modulation equations~\eqref{e:Whitham3D} 
is already written in vectorial, dimension-independent form,
with the only caveat that, with $N$ spatial dimensions, $\@q$ and $\@p$ have $N-1$ components.
Moreover, all the steps of the derivation in section~\ref{s:3d} are written in a way that generalizes to any number of spatial dimensions.
Indeed, one can introduce spherical coordinates in $N$ spatial dimensions 
by generalizing~\eqref{e:khat3D} as 
$\^k_1 = \cos\varphi_1$, 
$\^k_2 = \sin\varphi_1\cos\varphi_2$, 
$\^k_3 = \sin\varphi_1\sin\varphi_2\cos\varphi_3$, 
etc., up to 
$\^k_{N-1} = \sin\varphi_1\cdots\sin\varphi_{N-2}\cos\varphi_{N-1}$
and 
$\^k_N = \sin\varphi_1\cdots\sin\varphi_{N-2}\sin\varphi_{N-1}$.
Then, one introduces $q_1,\dots,q_N$ via the generalization of~\eqref{e:qdef3D}, 
namely, $q_1 = k_2/k_1$, $q_2 = k_3/k_1$, etc., up to $q_{N-1} = k_N/k_1$,
as well as  $p_1,\dots,p_N$ via the natural generalization of~\eqref{e:u2&u3}.
In this way, one obtains the generalization of~\eqref{e:gdef3D} as
$g = 1 + q_1^2 + \cdots + q_N^2 = \sec\varphi_1$, 
and all the calculations and equations in section~\ref{s:3d} remain valid
as long as one also redefines the operators $\bfnabla_\flat$ and $\@D_\flat$ accordingly.

We point out that, even though we have not done so explicitly in this work, 
it would be straightforward to obtain the reduction to the Whitham equations for the radial NLS equation from the 3DNLS-Whitham equations derived in section~\ref{s:3d} using spherical coordinates.
It would also be straightforward to write down explicitly the harmonic and soliton limits in three spatial dimensions,
as well as all of these corresponding limits in higher dimensions.

We should comment on the importance of the constraints $\bfnabla\wedge\@k=0$ and $\bfnabla\wedge\={\@u} = 0$. 
On one hand, these constraints play a key role in the derivation.
On the other hand, the final Whitham equations [e.g., \eqref{e:3D:diagonalW1to4}, \eqref{e:dqdt3D} and~\eqref{e:dpdt3D}]
do not automatically ensure that these constraints are satisfied,
only that their time derivative is (similar to \cite{ABW2017}).
Because of this, the compatibility between solutions of the Whitham system 
and modulated one-phase solutions of the NLS equation is not guaranteed a priori, and, 
similar to \cite{ABW2017}, 
one must give initial conditions that are compatible with the one-phase assumption.
These constraints are also likely to be related to the integrability properties of the system, 
as discussed below.

We emphasize that this work is foundational and that, similar to \cite{ABW2017}, 
the results presented here open up a number of interesting problems,
which are expected to lead to several further advances in the near future.
Specifically, we next mention and briefly discuss some of these possible of avenues for further research. 

One direction for future work is the derivation of the Whitham equations for the focusing NLS equation 
in three spatial dimensions.
We expect that this will be straightforward. 
Indeed, the Whitham equations in the two-dimensional focusing case were already written in \cite{ACR2021}
(although not in rotation-invariant form).
Once the derivation of the one-phase solutions of the NLS equation is done in dimension-invariant form,
as was the case in section~\ref{s:periodic}, 
the rest of the machinery presented in this work will carry over to the case of the focusing case in three and higher dimensions 
without significant changes.
Of course, as in the one-dimensional case, the resulting Whitham equations will be elliptic 
(i.e., the characteristic velocities will be complex), and therefore
require suitable interpretation of initial value problems; see \cite{BK2006,PHYSD1995v87p186,GurevichKrylov,PhysD236p44,PRA74p023623,SJAM59p2162} 
as well as \cite{PhysD333p1,ElHoefer} and references therein.
The Whitham equations for the NLS equation in one spatial dimension have also proved to be useful in some situations,
even in the focusing case 
\cite{B2018,BM2016,elgurevich},
so one can expect that those in two and three spatial dimensions will be useful as well.

Another important direction for future work is a study to determine whether the Whitham modulation system derived here, 
or any of its reductions, are completely integrable.
A notion of integrability for multidimensional systems was put forth in \cite{ferapontov2004,ferapontov2006},
based on the existence of infinitely many $N$-component reductions.
Of course, the NLS equation in more than one spatial dimension is not integrable,
and therefore one would have no reason to expect that the corresponding NLS-Whitham systems are.
Still, the reductions to one-dimensional NLS-Whitham equations are indeed integrable, and 
therefore it is a natural question whether there are other integrable reductions. 
In this regard, we should point out that, even for the KP equation (which is integrable), 
the original Whitham system derived in \cite{ABW2017} appears not to be integrable,
but its harmonic and soliton limits are \cite{BHM2020}.
Moreover, so are various less-trivial one-dimensional reductions 
beyond the obvious reduction to the Whitham system for the KdV equation, 
once one properly takes into account the analogue of the compatibility conditions
\eqref{e:conservationwaves1} and~\eqref{e:curlk}~\cite{BBH2022}.

Yet another interesting problem for future work is the issue of whether one can establish 
a precise relation between the 2DNLS-Whitham system and the KP-Whitham system.
It is well known that the 1DNLS-Whitham system admits a reduction to the KdV-Whitham system \cite{GurevichKrylov}.
It is also well known that the 2DNLS equation admits a reduction to the KP equation \cite{KuznetsovTuritsyn1988}.
A natural question is therefore whether the 2DNLS-Whitham system admits a reduction to the KP-Whitham system.
It is straightforward to see that,
if one considers the same reduction as in \cite{GurevichKrylov},
the PDEs for $r_1,\dots,r_4$ in the 2DNLS-Whitham system naturally reduce to those for $r_1,\dots,r_3$
in the KP-Whitham system. 
The PDE for $q$ also reduces to the corresponding equation in the KP-Whitham system, 
since it just comes from the second component of the conservation of waves equation
in both systems.
The open question, however, is how one can obtain a PDE for~$p$ that does not contain a time derivative,
as prescribed in the KP-Whitham system.  

Finally, and most importantly from a practical point of view, an
obvious opportunity for future work will be the use of the modulation
equations derived here to characterize the dynamical behavior in
physically significant scenarios.
One important application is to the
description of dispersive shock waves (DSWs)
\cite{GurevichKrylov,PhysD333p1}.
Some of the earliest experiments on
DSWs in nonlinear optics and Bose-Einstein condensates (BECs)---where
the defocusing NLS equation is an excellent model---involved
inherently multidimensional nonlinear wave propagation
\cite{Dutton2001,Simula2005,PRA74p023623,Wan2007}. One intriguing
feature, observed in both BEC and optics \cite{PRA74p023623,Wan2007},
is the coherent propagation of multidimensional DSWs with stable
ring/spherical and elliptical/ellipsoidal patterns.
These
observations are at odds with the known transverse instability of
planar cnoidal wave solutions of \eqref{e:nls} \cite{Thelwell2006}.
Further analysis of the 2D and 3DNLS-Whitham modulation equations may
provide some analytical insight in this.
Moreover, BECs are
three-dimensional, so the (3+1)-dimensional modulation equations
derived here are needed to describe large amplitude matter waves.
Three-dimensional effects have been shown to be decisive in some BEC
DSW experiments \cite{Chang2008,Mossman2018}.

Various applications of the Whitham equations for the focusing and
defocusing NLS equations in one spatial dimension were already
mentioned above.
We should also note that, while the full modulation
system composed of equations \eqref{e:3D:diagonalW1to4},
\eqref{e:dqdt3D} and~\eqref{e:dpdt3D} might appear complicated, even
its reductions can be useful in this regard.
For example, of
particular interest from an applicative point of view are the harmonic
and soliton limits.
In the one-dimensional case, soliton modulation
theory and its applications were studied for the KdV equation in
\cite{PRL120p144101} and for the defocusing NLS equation in
\cite{Sprenger2018}, while the harmonic limit of the Whitham equations
for the KdV equation was studied in \cite{Congy2019}.
Similarly, the
harmonic and soliton limits of the Whitham equations for the KP
equation, which were derived and analyzed in \cite{ABW2017,BHM2020}
have found concrete applications in
\cite{Ryskamp2021,Ryskamp2021a,Ryskamp2022}.
These reductions
analytically describe the evolution of a soliton or linear waves in
the presence of the slowly varying mean field $\=\rho$, $\=\@u$.
Obtaining these modulation equations using multiple scales and a
soliton ansatz is quite tedious and, to our knowledge, has apparently
only been carried out for the KdV equation in \cite{Grimshaw1979}.
We
believe that, like with Whitham equations for the KP equation
\cite{ABW2017}, the modulation equations derived in this work will
prove to be an effective tool to study several physically significant
problems.
The soliton limit should prove to be particularly important
in this respect, similar to the KP equation
\cite{Ryskamp2021,Ryskamp2022,Ryskamp2021a}.

We hope that the results of this work and the present discussion will
provide a stimulus for several further studies on these and related
problems.

\subsection*{\bf Acknowledgments}

We thank Alexandr Chernyavskiy and Dmitri Kireyev for many useful discussions on topics related to this work.
This research was partially supported by the National Science Foundation under grant numbers DMS-1816934 and DMS-2009487.

\section*{Appendix}
\setcounter{section}1
\def\thesection{\Alph{section}}
\def\theequation{\Alph{section}.\arabic{equation}}
\def\numparts{\refstepcounter{equation}%
     \setcounter{eqnval}{\value{equation}}%
     \setcounter{equation}{0}%
     \def\theequation{\Alph{section}.\arabic{eqnval}{\it\alph{equation}}}}
\def\endnumparts{\def\theequation{\Alph{section}.\arabic{equation}}%
     \setcounter{equation}{\value{eqnval}}}

\subsection{Direct derivation of the periodic solutions of the NLS equation}
\label{s:directperiodic}
Here we derive the periodic solutions of the NLS equation in an arbitrary number of dimensions directly, 
without using the hydrodynamic system.
We start with the one-phase ansatz
\vspace*{-0.2ex}
\be
\psi(\@x,t) = \sqrt{\rho(z/\epsilon)}\,\e^{i\Phi(z/\epsilon)}\,,
\label{e:qansatz0}
\ee
where, as before, the ``fast variable'' is $Z = \@k\cdot\@x - \omega t$.
Substituting~\eqref{e:qansatz0} into \eqref{e:nls} and separating into real and imaginary parts yields respectively:
\bse
\bea
(\sqrt{\rho})''- \sqrt{\rho}(\Phi')^2 + \frac\omega{\|\@k\|^2} \sqrt{\rho}\,\Phi' - \frac{2}{\|\@k\|^2}\rho^{3/2} = 0 \,,
\label{e:nlsrhophireal}
\\
\sqrt{\rho}\,\Phi'' + \Big(2\Phi' - \frac\omega{\|\@k\|^2} \Big)(\sqrt{\rho})' = 0\,,
\label{e:nlsrhophiimag}
\eea
\ese
where, for brevity in this section, we denote $a = \|\@k\|^2$.
Integrating \eqref{e:nlsrhophiimag} yields $\Phi'$ up to an integration constant $J$
\be
\Phi' = \frac{J}{\|\@k\|\,\rho}+ \frac\omega{2a}\,.
\label{e:phaserelation}
\ee
Substituting the phase relation~\eqref{e:phaserelation}, the real part
\eqref{e:nlsrhophireal} reduces to: 
\bea
(\sqrt{\rho})''- \frac{J^2}{a\,\rho^{3/2}} +\Big(\frac\omega{2a}\Big)^2 \sqrt{\rho} - \frac{2}{a} \rho^{3/2} = 0.
\eea
Multiplying by $2(\sqrt{\rho})'$ and integrating with respect to $Z$ and letting $f = \rho$ yields: 
\bea
\label{e:fode}
(f')^2 = \frac{4}{a} f^3 -4 \Big(\frac\omega{2a}\Big)^2 f^2 + 4c_1f - \frac{4J^2}{a}\,.
\eea
By substituting $f(Z) = A + B y^2(Z)$, we get the following ODE for $y$:
\bea
\displaystyle
\fl
(y')^2 = \frac{1}{B^2} \Big[\frac{A^3}{a}- A^2 \Big(\frac\omega{2a}\Big)^2 + Ac_1 - \frac{J^2}{a}\Big]\frac{1}{y^2} + \frac{1}{B} \Big[\frac{3A^2}{a} - 2A \Big(\frac\omega{2a}\Big)^2 + c_1\Big]
+ \Big[ \frac{3A}{a} - \Big(\frac\omega{2a}\Big)^2\Big]y^2 + \frac{B}{a}y^4.
\label{e:yODE0}
\eea
Now recall that the Jacobian elliptic sine $y(Z) = \sn(cZ|m)$ solves the ODE $(y'/c)^2 = (1 - y^2)(1-m y^2)$.
By requiring that \eqref{e:yODE0} matches the ODE for the elliptic sine, one then obtains~\eqref{e:rho},
with $B = 4m \|\@k\|^2K_m^2$ as before,
and with
\bse
\bea
\label{e:BDJ}
J^2 = 4 a K_m^2 A\bigg(1+\frac{A}{4 K_m^2 a}\bigg)(A+4 m K_m^2 a)\,,
\label{e:J20}
\\
\label{e:T0}
\Big(\frac\omega{2a}\Big)^2 = 4 K_m^2 (1 + m) + \frac{3A}{a} \,,\quad
c_1 = \frac1a\,\big[\big({4 m K_m^2 a} + A\big)(4 K_m^2 a + 2A) + A\big({4 K_m^2 a} + A\big)\big]\,.
\eea
\ese
Similar to section~\ref{s:periodic}, we write the ODE \eqref{e:fode} as
$(f')^2 = P_3(f)$,
where 
\be
P_3(f) = \frac{4}{a}\Big[f^3 - a \Big(\frac\omega{2a}\Big)^2 f^2 + c_1\,a\,f - J^2 \Big] 
  = \frac{4}{a} ( f - \lambda_1 )( f - \lambda_2 )( f - \lambda_3 )\,,
\ee
with $\lambda_1,\dots,\lambda_3$ given by~\eqref{e:lambda123_fromAam}.
Note that the requirements $a\ge0$ and $0\le m\le 1$ again immediately imply
\eqref{e:lambdaconstraint}.
The symmetric polynomials defined by $\lambda_1$, $\lambda_2$, and $\lambda_3$ are related to the above constants as 
\be
\label{e:symmetricpolys0}
e_1 = \lambda_1 + \lambda_2 + \lambda_3 = {\omega^2}/{4a}\,,\quad
e_2 = \lambda_1 \lambda_2 + \lambda_2\lambda_3 + \lambda_3\lambda_1 = c_1\,a\,,\quad
e_3 = \lambda_1 \lambda_2 \lambda_3 = J^2.
\ee
which also allow one to recover $A$, $a$, and $m$  when $\lambda_1$,
$\lambda_2$, and $\lambda_3$ are known.
The above solution contains $N+2$ independent parameters: $A$, $m$ and $\@k$
[since $J$ and $\omega$ are determined by~\eqref{e:BDJ}, \eqref{e:T0}].
Next we employ the Galilean invariance of the NLS equation to apply a Galilean boost and thereby 
obtain the more general family of solutions
\bse
\be
\label{e:generalperiodicsolution}
\~\psi(\@x,t) = \psi(\@x-2\@v t,t)\,\e^{i(\@v\cdot\@x-\|\@v\|^2t)/\epsilon}
  = \sqrt{\rho(\~z/\epsilon)}\,\e^{i\~{\Phi}(\~z/\epsilon,\@x,t)}\,,
\ee
where 
$\~z = \@k\cdot\@x - \~\omega t\,$, 
with 
$\~\omega = \omega + 2 \@k\cdot\@v$,
and where 
\be
\~{\Phi}(\~z/\epsilon,\@x,t) = \Phi(\~z/\epsilon) + (\@v\cdot\@x - \|\@v\|^2t)/\epsilon\,.
\ee
\ese
The transformation adds the $N$ new independent parameters $v_1,\dots,v_N$.
Therefore, the periodic solution of the NLS equation~\eqref{e:nls} in $N$ spatial dimensions
contains $2N+2$ independent real parameters: $A$, $m$, $\@k$ and $\@v$, as expected.

\subsection{Calculation of the solution amplitude and second frequency and simplification of certain terms}
\label{a:ODE}

Here we give a few additional details on the calculation of the fluid density.
Starting from~\eqref{e:hleadingh2}, using~\eqref{e:uz} and 
simplifying the resulting ODE, one has
\be
a\rho''' + a\rho'\frac{(\rho')^2 - 2 \rho\rho''}{\rho^2} - 4 \rho \rho' + \frac{4J^2\rho'}{\rho^2} = 0\,,
\ee
where $a = \|\@k\|^2$ for brevity.
Integrating w.r.t.\ $Z$ yields
\be
a\rho'' - a \frac{(\rho')^2}{\rho} - 2\rho^2 + 2c_1 - \frac{4J^2}{\rho} = 0\,,
\label{e:rho''}
\ee
where $c_1$ is an arbitrary integration constant.
Multiplying~\eqref{e:rho''} by $2\rho'/\rho^2$ and integrating with respect to $Z$ again yields
\bea
\label{e:ode}
a(\rho')^2 = 4\rho^3 - 4c_2\rho^2 + 4c_1\rho - 4J^2\,,
\eea
with $c_2$ another arbitrary integration constant.
Letting $\rho(Z) = A + B y^2(Z)$ yields the following ODE:
\bea
(y')^2 = \frac{1}{B^2a} (A^3 - A^2 c_2 + A c_1  - J^2) \frac{1}{y^2} + \frac{1}{Ba} (3 A^2 - 2A c_2 + c_1)
+ \Big(\frac{3A}{a} - \frac{c_2}{a}\Big)y^2 + \frac{B}{a}y^4.
\label{e:yODE}
\eea
Now recall that the Jacobi elliptic sine $y(Z) = \sn(cz|m)$ solves the ODE $(y'/c)^2 = (1 - y^2)(1-m y^2)$.
By requiring that \eqref{e:yODE} matches the ODE for the elliptic sine, one obtains~\eqref{e:rho},
with the coefficients as in~\eqref{e:BDJ}.

Next we obtain \eqref{e:mu}, which determines the frequency $\mu$ of the second phase.
As mentioned in section~\ref{s:hydrodynamic}, to this end one can use the undifferentiated version of \eqref{e:hleadingh2} 
[obtained from the real part of~\eqref{e:nls}
using~\eqref{e:madelungansatz} and~\eqref{e:twophaseanstaz}], 
which is
\vspace*{-0.6ex}
\be
\label{e:undiffleading}
-\omega \phi' + 2(\@k\cdot\={\@u})\,\phi' + a(\phi')^2 + 2\rho -\mu + \|\={\@u}\|^2 - \frac{a}4 \Big((\ln \rho)'' + \frac{\rho''}{\rho}\Big) = 0\,. 
\ee
Differentiating~\eqref{e:undiffleading} w.r.t.\ $x$ and $y$ and collecting leading-order terms yields \eqref{e:hleadingh2}.
However, \eqref{e:undiffleading} allows us to determine $\mu$ in a more straightforward manner.
Indeed, substituting~\eqref{e:hphaserelation1} into equation \eqref{e:undiffleading} and simplifying yields,
\bse
\be
2a\rho'' - a \frac{(\rho')^2}{\rho} - 8\rho^2 + C \rho - \frac{4J^2}{\rho} = 0\,,
\label{e:rho''2}
\ee
where 
\be
C = 4\mu -4\big(\|\=u\|^2 - (J\,\overline{\rho^{-1}}/g^{1/2})^2 \big)
\label{a:Cdef}
\ee
\ese
Multiplying~\eqref{e:rho''2} by $\rho'/\rho$ and integrating with respect to $Z$ yields
\bea
\label{e:undiffode}
a(\rho')^2 = 4\rho^3 - C\rho^2 + 4c_3\rho - 4J^2\,,
\eea
with an arbitrary integration constant  $c_3$.
Comparing the coefficients in \eqref{e:ode} and \eqref{e:undiffode} we have $C = 4 c_2$ [as well as $c_1 = c_3$],
which, when inserted in~\eqref{a:Cdef}, finally yields~\eqref{e:mu} for~$\mu$.

Finally, we provide further details on how to simplify the modulation equations~\eqref{e:averagedconsevationlaws}
and in particular on how to obtain~\eqref{e:2pw6a}.
The averaged conservation of momentum equation~\eqref{e:acl6},
when written in terms of $\lambda_1,\dots,\lambda_3$ and~$\@U$,
is 
\bea
(J\,\hat{\@k} + \=\rho \@U)_t +
\nabla(\overline{\rho^2}) + 
  \nabla\cdot \bigg[ 2\=\rho\@U\otimes \@U  + 2J\,(\hat{\@k}\otimes \@U + \@U\otimes\hat{\@k})
   + \bigg(\bigg(\overline{\frac{(\rho')^2}{2\rho}}\bigg)
   + 2\frac{J^2 \overline{\rho^{-1}}}{\|\@k\|^2} \bigg)\@k\bigg) \otimes \@k \bigg]
  = \@0\,.
\nonumber\\
\label{e:2pw6}
\eea
Notice that averages containing $\rho_z$ can be evaluated by recalling that $\rho(Z)$ satisfies the ODE~\eqref{e:drhodz}.
Differentiating~\eqref{e:drhodz} and using the definition of the symmetric polynomials yields
\bse
\be
\frac{\|\@k\|^2}{2}\rho'' = 3\rho^2 -2e_1 \rho + e_2\,,
\label{e:rho_zz}
\ee
and averaging over the fast variable $Z$ gives
\be 
\overline{\rho^2} = \frac{2e_1 \overline{\rho} - e_2}{3}\,.
\label{e:barrho^2}
\ee
Reordering the ODE~\eqref{e:cubicpolynomial} gives us
\be
\frac{\|\@k\|^2 \rho_2^2 + 4 J^2}{\rho} = 4\rho^2 - 4e_1\rho + 4 e_2\,.
\ee
Averaging again over the fast variable $Z$ and using \eqref{e:barrho^2} yields
\be
\frac{\|\@k\|^2}{4}\Big(\overline{\big(\frac{(\rho')^2}{\rho}\big)} + \frac{4J^2}{\|\@k\|^2}\overline{\rho^{-1}}\Big) = \frac13{(2e_2 - e_1\overline{\rho})}\,.
\label{e:replacebarrho-1}
\ee
\ese
Finally, using~\eqref{e:replacebarrho-1}, equation~\eqref{e:2pw6} yields \eqref{e:2pw6a}.

\subsection{Detailed steps in the derivation of the 3DNLS-Whitham system}
\label{a:3Dsimplification}

We begin by expressing the modulation equations in terms of convective derivatives.
Using~\eqref{e:qidentity} one can see that
\be
D_y(q_2) = q_{2,y} - q_1 q_{2,x} = \frac{D_z(k_1q_1)}{k_1} - q_1 \frac{D_z k_1}{k_1} = D_z(q_1)\,,
\label{e:qrelation2}
\ee
which proves~\eqref{e:qrelation}.
Moreover, using~\eqref{e:curlu} and the fact that $\@p = \={\@u}_\flat - \=u_1 \@q$, one has
\bse
\bea
D_y(p_2) &= (\={u_3} - q_2 \={u_1})_y - q_1 (\={u_3} - q_2 \={u_1})_x = \=u_{3,y} - q_1 \=u_{3,x} - q_2 \=u_{1,y} + q_1 q_2 \=u_{1,x} - \=u_1 D_y(q_2) \,,
\nonumber
\\
&= \=u_{2,z} - q_1 \=u_{1,z} - q_2 \=u_{2,x} + q_1 q_2 \=u_{1,x} - \=u_1 D_z(q_1) = D_z(p_1)  \,.
\eea
\label{e:prelation2}
\ese
which yields~\eqref{e:prelation}.

Using the identity \eqref{e:qrelation} and straightforward algebra,
we can rewrite~\eqref{e:dqdt3D} and~\eqref{e:dpdt3D} as
\bse
\bea
\label{e:Dq}
D_t \@q + 2g \@D_\flat U_1 + 2q_1 (U_1 \@D_\flat q_1 + \@D_\flat p_1)  + 2q_2 (U_1 \@D_\flat q_2 + \@D_\flat p_2) = 0\,,
\\
\label{e:Dp}
D_t\@p - 2q_1 U_1 D_y \@p - 2q_2 U_1 D_z \@p + \@D_\flat (g(\~e_1 - U_1^2)) = 0\,,
\eea
\ese
where $D_t$ is as in~\eqref{e:Dx}.

Next, we express the first conservation of waves equation in convective derivative form.
Recalling equations~\eqref{e:k1_t} and using \eqref{e:qidentity} one can obtain the following,
\be
\label{e:auxW1}
\frac{D_t k_1}{k_1} + 2(U_1)_x + 2 \@q \cdot (\@U_\flat)_x  = 0\,.
\ee
Simplifying further we have~\eqref{e:DK1}, 
with $D_x$ as in~\eqref{e:Dx} and with 
\be
W_1 = U_1 (\|\@q\|^2)_x  - 2\@q \cdot \@D_\flat U_1 + 2 \@q \cdot \@p_x\,.
\ee
Moreover, using~\eqref{e:Dq} one can simplify $W_1$ further and obtain~\eqref{e:gW_1}.

Next, it can be easily seen that~\eqref{e:(u+s)_t} becomes
\bea
D_t (U_1+J\,\overline{\rho^{-1}}/g^{1/2}) + g (\~e_1)_x + 2gJ\,\overline{\rho^{-1}}/g^{1/2} (U_1)_x + 2U_1 \|\@q\|^2 (J\,\overline{\rho^{-1}}/g^{1/2}+U_1)_x
\nonumber
\\\kern2em
- 2(U_1 \@q + \@p) \cdot \bfnabla_\flat (J\,\overline{\rho^{-1}}/g^{1/2}+U_1)
+ 2\@p \cdot \big((J\,\overline{\rho^{-1}}/g^{1/2}+U_1)\@q\big)_x 
\nonumber
\\\kern2em
+ (\~e_1 + U_1^2 + 2J\,\overline{\rho^{-1}}/g^{1/2}U_1) \|\@q\|^2_x + 2 (\@p + U_1\@q +J\,\overline{\rho^{-1}}/g^{1/2}\@q) \cdot \@p_x = 0\,.
\label{e:auxW2}
\eea
As a direct consequence of equation \eqref{e:2pw4} we obtain
\be
\label{e:auxcurlfree}
\@p_x = \bfnabla_\flat (U_1 + J\,\overline{\rho^{-1}}/g^{1/2}) - \big((U_1+J\,\overline{\rho^{-1}}/g^{1/2})\@q\big)_x\,.
\ee
Using the above relation for $\@p_x$ and eliminating $(U_1)_x$ with the help of~\eqref{e:auxW1}, 
one can obtain the simplified second wave conservation equation~\eqref{e:D(U+S)} in terms of convective derivatives,
where
\bea
W_2 = \txtfrac12\big[(\~e_1 + U_1^2) \|\@q\|_x + 2U_1 \@q \cdot \big(2 (\@U_\flat)_x - \bfnabla_\flat U_1 + J\,\overline{\rho^{-1}}/g^{1/2} \@q_x - \@D_\flat S \big)
\nonumber\\\kern2em
- \@q \cdot \@D_\flat (\~e_1 + U_1^2) - (U_1^2 \|\@q\|^2)_x\Big].
\eea
Equation \eqref{e:auxcurlfree} yields
\be
J\,\overline{\rho^{-1}}/g^{1/2} \@q_x - \@D_\flat\big(J\,\overline{\rho^{-1}}/g^{1/2}\big) = \@D_\flat U_1 - \@p_x - U_1 \@q_x\,.
\ee
Using the above relation along with equations \eqref{e:Dq}, \eqref{e:Dp}, \eqref{e:prelation}, \eqref{e:qrelation}
and some tedious but straightforward algebra, one obtains~\eqref{e:gW_2}.

To express the averaged mass equation \eqref{e:rhobar_t} in terms of convective derivatives, first we replace $\@M$ using \eqref{e:M} and obtain:
\be
D_t (\overline\rho) + 2 \overline\rho \big( (U_1)_x + \bfnabla_\flat \cdot \@U_\flat \big) + 2 \big( (g \~J)_x + \bfnabla_\flat \cdot (g\~J \@q)\big) = 0\,,
\ee
Simplifying further we obtain
\be
\label{e:auxW3}
D_t (\overline\rho) + 2 \overline\rho D_x U_1 + 2 D_x (g \~J) + 2 g M_1 (\bfnabla_\flat \cdot \@q) + 2 \overline\rho (\bfnabla_\flat \cdot \@p) =0\,.
\ee
To rewrite~\eqref{e:auxW3} in simpler form, 
we consider the combination \eqref{e:auxW3} $-~g (\overline\rho/g)$\eqref{e:DK1}, which yields~\eqref{e:W3}.

Finally, we consider the first component of the averaged momentum equation \eqref{e:M1_t}, using a similar approach as before one can rewrite it as follows:
\bea
\label{e:DtM1}
D_t(gM_1) + 2g (M_1 + \~J) D_x U_1 + 2U_1 D_x(g\~J) + D_x(g\~e_2) - \@q \cdot \@D_\flat (g\, (\overline\rho^2/g^2))
\nonumber
\\
\kern7em
+ 2(\overline\rho^2/g^2) g (\@q \cdot \@q_x)
+ g \big(\~e_2 - (\overline\rho^2/g^2) + 2U_1(M_1 + \~J)\big) \bfnabla_\flat \cdot \@q + 2M_1 g \bfnabla_\flat \cdot \@p =0\,.
\eea
Next, taking the combination \eqref{e:DtM1} $-~U_1$\eqref{e:auxW3} yields
\bea
\label{e:aux1W4}
g (\overline\rho/g) D_t U_1 + D_t (g \~J) + 4g\~J D_x U_1 + D_x (g\~e_2) - \@q \cdot \@D_\flat(g (\overline\rho^2/g^2))
\nonumber
\\
\kern9em
+ 2 g (\overline\rho^2/g^2) (\@q \cdot \@q_x)  
+ g (\~e_2 - (\overline\rho^2/g^2) - 2U_1\~J) + 2g \~J \bfnabla_\flat \cdot \@p =0\,.
\eea
To simplify this equation more we consider the combination \eqref{e:M2&M3}$-$\eqref{e:M1_t}$\@q -$\eqref{e:rhobar_t}$\@p$ and obtain the following vector equation:
\bea
\label{e:aux2W4}
M_1 D_t \@q + (\overline\rho/g) D_t \@p + (2U_1\~J + \~e_2) D_x \@q + 2\~J D_x \@p
\nonumber\\\kern10em
  - (\overline\rho^2/g^2) D_x \@q + g \@D_\flat ((\overline\rho^2/g^2)) + 2(\overline\rho^2/g^2) \@D_\flat g = 0\,.
\eea
Finally, we consider the combination \eqref{e:aux1W4}$-2g\~J\,$\eqref{e:DK1}$+\@q \cdot\,$\eqref{e:aux2W4}.
Using \eqref{e:qrelation} and after extensive simplifications, we obtain~\eqref{e:W4}.

Next we show that, using the transformation to Riemann-type variables~\eqref{e:lambda123_to_r1234}
\eqref{e:DK1}, \eqref{e:D(U+S)}, \eqref{e:W3} and \eqref{e:W4} 
yield~\eqref{e:waveW1234}.
Note first that, using~\eqref{e:lambda123_to_r1234}, we have the
following identities:
\bse
\label{e:symminriemann}
\bea
&U_1 = \txtfrac 12 s_1\,,
\label{e:Us1}
\\
&\~e_1 = s_2 - \txtfrac14 s_1^2\,,
\\
&\~e_2 = s_4 + \txtfrac1{16}\,[ - 16 s_3 s_1 - 4 s_2^2 + 8 s_1^2 s_2 - s_1 ^4 ]\,,
\\
&
\~J = \txtfrac13 s_3 - \txtfrac1{24}\,s_1(6s_2 - s_1^2)\,,
\label{e:e3identity}
\eea
\ese
[where again the $s_n$ are as in~\eqref{e:sdef}],
which allow us to express $\~e_1,\dots,\~e_3$ in terms of the Riemann invariants via $s_1,\dots,s_4$.
The identity~\eqref{e:e3identity} is especially important, since it allows us to eliminate square roots from the
modulation equations.
Recall that~\eqref{e:symmetricpolys} only determines $J^2$, and 
$J = \sigma(\lambda_1\lambda_2\lambda_3)^{1/2}$,
with $\sigma=\pm1$.
On the other hand, \eqref{e:lambda123_to_r1234} 
yields
$\sigma\lambda_1^{1/2} = \half\,\sqrt{g}\,(r_1 - r_2 - r_3 + r_4)$,
where the sign~$\sigma$ here is needed because
but one needs $\lambda_3\ge\lambda_2\ge\lambda_1\ge0$
(cf.\ section~\ref{s:periodic}),
but $r_1 - r_2 - r_3 + r_4$ 
can be either positive or negative depending on the relative magnitude of $r_1,\dots,r_4$.
(In contrast, no ambiguity arises for $\lambda_2^{1/2}$ and $\lambda_3^{1/2}$ when $r_1,\dots,r_4$ are well-ordered.)
One can verify that, with this choices, 
the sign of both the left-hand side and right-hand side of \eqref{e:e3identity} equal~$\sigma$.
The following formulae are also useful:
\bse
\bea
\fl
k_1 = \frac{\sqrt{(r_4-r_2)(r_3-r_1)}}{2\K}\,,
\label{e:k1inr}
\\
\fl
\bigg(\partialderiv {k_1}{r_1},\dots,\partialderiv {k_1}{r_4}\bigg)^T = \frac{\sqrt{(r_4-r_2)(r_3-r_1)}}{4K_m^2}\,
\bigg(
\frac{(r_1-r_4) K_m + (r_4-r_2)E_m}{(r_2-r_1)(r_4-r_1)},
\nonumber
\\
\frac{(r_3-r_2) K_m + (r_1-r_3) E_m}{(r_2-r_1)(r_3-r_2)},
\frac{(r_3-r_2)K_m + (r_2-r_4) E_m}{(r_3-r_2)(r_3-r_4)},
\frac{(r_4-r_1) K_m + (r_1-r_3)E_m}{(r_4-r_1)(r_4-r_3)}\bigg)^T\,,
\label{e:gradkr}
\eea
\ese

We are now ready to present the final steps of the derivation.
We begin by deriving~\eqref{e:DK1}, which is the simplest of the four equations. In this case we simply need to express 
$D_t k_1$ in terms of the Riemann invariants, i.e.,
\vspace*{-1ex}
\be
D_t k = \sum_{j=1}^4 \frac{\partial k_1}{\partial r_j}\,D_t r_j\,,
\ee
which immediately yields~\eqref{e:waveW1},
with $W_1$ as in~\eqref{e:gW_1}.
Next, equation \eqref{e:D(U+S)} simplifies due to the identity
\be
\label{e:identityW2}
D_t (U_1+J\,\overline{\rho^{-1}}/g^{1/2})  + (U_1-J\,\overline{\rho^{-1}}/g^{1/2})\,\frac{D_t k_1}{k_1}  = \frac 2{k_1} \sum_{j=1}^4 r_j\partialderiv {k_1}{r_j} D_t r_j\,,
\ee
and takes the form of~\eqref{e:waveW2}, 
with $W_2$ as in~\eqref{e:gW_2}.
Next, taking the combination~\eqref{e:W3}/2 + $gU_1/2\times $\eqref{e:D(U+S)} + $g(s_2-2U_1^2)/4\times $\eqref{e:DK1} and 
using identities~\eqref{e:symminriemann} and \eqref{e:identityW2}, yields~\eqref{e:massW3},
where $W_3$ is as in~\eqref{e:gW3}.
Finally, considering the linear combination \eqref{e:W4}/2 + $3U_1/2\times$\eqref{e:W3}  $g(s_2 + 2U_1^2)/4\times$\eqref{e:D(U+S)} + $g(3\~J + U_1s_2 -2U_1^3)/2\times$\eqref{e:DK1}, and using identities~\eqref{e:symminriemann}, \eqref{e:identityW2} again, and after some tedious algebra, one finds~\eqref{e:momentumW4},
with $W_4$ as in~\eqref{e:gW4}.

Our last task is to show that the compatibility relations $\bfnabla\times\@k = \bfnabla\times\={\@u} = \@0$, 
when written in terms of the Riemann-type variables $\@r = (r_1,\dots,r_4)^T$ as well as $\@q$ and $\@p$,
yield~\eqref{e:curlfreeRiemann}.
To this end, we first use the definition of $\@q$ as in \eqref{e:qdef3D} along with the compatibility condition $\bfnabla\times\@k = 0$. It can be easily seen that
\be
k_1 \@q_x = (\@k_\flat)_x - k_{1,x}\, \@q = \bfnabla_\flat k_1 - k_{1,x}\, \@q = \@D_\flat k_1\,
\ee
(cf.\ the third equation in~\eqref{e:conservationwaves1p&q}),
which yields the first half of~\eqref{e:curlfreeRiemann}. 
Next, using the compatibility condition $\bfnabla\times\={\@u} = 0$ with the definition of $\=u_1$ as in \eqref{e:u1&U1} 
one can derive \eqref{e:p_x}, namely,
\be
{\@p}_x = \@D_\flat\big( U_1 + J\,\overline{\rho^{-1}}/g^{1/2}\big) + \big(U_1 - J\,\overline{\rho^{-1}}/g^{1/2}\big)\frac{\@D_\flat k_1}{k_1} - 2U_1 \frac{\@D_\flat k_1}{k_1}\,.
\ee
Using the identity \eqref{e:identityW2}, 
one then obtains the second half of~\eqref{e:curlfreeRiemann}.

\section*{References}

\def\journal#1#2{\textit{#1}\unskip~\textbf{\ignorespaces #2}}
\def\v#1{\textbf{#1}}

\def\reftitle#1{``#1''}
\let\title=\reftitle

\end{document}